\documentclass{aa}  

\usepackage{graphicx}
\usepackage{txfonts}
\usepackage{lipsum}
\usepackage{subcaption}         
\usepackage{lscape}             
\usepackage{placeins}           

\begin{document}

   \title{A complete survey of filaments in Cygnus X}

   \subtitle{}

   \author{Yingxi Li\inst{1,2}
        \and Keping Qiu\inst{1,2}}

   \institute{School of Astronomy and Space Science, Nanjing University, 163 Xianlin Avenue, Nanjing 210023, People's Republic of China\\
             \email{ kpqiu@nju.edu.cn}
            \and Key Laboratory of Modern Astronomy and Astrophysics (Nanjing University), Ministry of Education, Nanjing 210023, P.R.China\\ }

   \date{submitted 30 October 2025}

  \abstract
   {Filamentary gas structures are widely observed in molecular clouds and are suggested to play a key role in the star formation processes. However, existing observations of molecular filaments are still biased toward case studies and small samples, and a large sample study from a single giant molecular cloud is lacking.}
   {We aim to carry out a complete census of filaments in Cygnus X and obtain their physical properties, relations with dense cores, magnetic field (B field), and HII regions.}
   {We extracted 2633 filaments and 6551 cores from the column-density map of Cygnus X, using the most updated \emph{getsf} algorithm. We then investigated the mass functions for the cores on and separate from the filaments, compared the orientations of the filaments with that of the B field obtained with the \emph{Planck} data, derived the radial column-density profiles of the filaments close to HII regions, and calculated the distances between the identified young stellar objects and filament spines. We also re-extracted filaments at the resolution of the \emph{Planck} 353 GHz dust-emission map to study their relationship with the B field.} 
   {The filaments in Cygnus X have a typical width of 0.5 pc. More than 93$\%$ of high-mass cores  ($\ge$ 20 $M_{\sun}$) are located on filaments. The core mass function constructed from the cores on the filaments (onCMF) shows a power law in the high-mass-end ($>10\,M_{\odot}$) with a slope of $-2.30$, whereas the high-mass end of the core mass function (CMF) derived from the cores outside the filaments (outCMF) has a much steeper power-law distribution with a slope of $-2.83$. The core mass corresponding to the peak of onCMF is much less than the Bonner-Ebert mass, but that corresponding to the peak of outCMF is well comparable to the Bonner-Ebert mass. Filaments re-extracted from the column-density map smoothed to an angular resolution identical to the \emph{Planck} 353 GHz map are mostly perpendicular to the B field, except that those of the lowest column densities are parallel to the B field. The transition from parallel to perpendicular occurs at a column-density equivalent to $A_V=10$ mag. Most prominent filamentary gas structures and high-mass cores appear to be preferentially located along the boundaries of HII regions or at the intersections of multiple HII regions. The filaments close to the HII region boundaries show a steeper column-density profile on the side toward the HII region compared to that on the opposite side.}
   {The formation of more massive cores has a stronger dependence on filaments, and the latter may provide a mass reservoir for the former to grow in mass via accretion. The B field plays a crucial role in filament formation, and the type-O mode where filaments form at the tip of converging flows along an oblique MHD shock front may be prevalent in Cygnus X. In this context, expanding HII regions in the complex induces shocks that compress the surrounding gas, creating inhomogeneity and dense clumps, and forming filaments. The global picture of filaments, cores, and star formation in Cygnus X is apparently consistent with the bubble-filament paradigm proposed in the literature.}
   

   \maketitle

\section{Introduction}
Gas filaments, including large giant filaments referred to as "Galactic bones" or "Galactic giant filaments" \citep[e.g.][]{Li2013,Goodman2014,Zucker2015} and smaller filamentary structures in molecular clouds \citep[e.g.][]{Arzoumanian2011, Palmeirim2013,Arzoumanian2021,Andre2025} are commonly seen in the interstellar medium \citep[ISM; e.g.][]{Hacar2023}. In particular, several \emph{Herschel} imaging surveys, such as the \emph{Herschel} Gould Belt Survey (HGBS) \citep[e.g.][]{Menhchikov2010, MivilleDeschenes2010, DiFrancesco2020} and the Herschel infrared Galactic Plane Survey (Hi-GAL) \citep{Molinari2010, Schisano2014, Kumar2020}, have revealed ubiquitous networks of filaments based on dust-continuum emission. These observations have established a significant correlation between the filaments and star-forming cores \citep{Andre2014, Andre2017, Hennebelle2019}. Studies, such as \citet{Konyves2015} in Aquila have revealed that approximately 75 percent of pre-stellar cores in that region are located within the supercritical filaments with masses per unit length of $M_{\rm line}$ $\textgreater$ $M_{\rm line,crit}$, where $M_{\rm line,crit}$ = 2$c_{\rm s}^{2}$/G $\sim$ 16$M_{\sun}$/pc is the critical mass per unit length of nearly isothermal, long cylinders at T $\sim$ 10 K. These studies suggest a scenario where large filaments may undergo gravitational fragmentation, leading to the formation of dense star-forming cores \citep{Hacar2023}. However, it is still not clear how filaments influence the shape of the core mass function (CMF). Cores located on the filaments tend to have higher surface densities, sometimes exceeding the theoretical threshold of high-mass star formation proposed in \citet{Krumholz2009}. This implies a critical role played by filaments in facilitating the growth of high-mass star-forming cores.

The HGBS survey suggests a two-stage process for core formation from molecular clouds \citep{Andre2014, Andre2019, DiFrancesco2020, Ladjelate2020, Shimajiri2023}. In the first stage, the dissipation of kinetic energy in large-scale magneto-hydrodynamic (MHD) flows leads to the generation of web-like filamentary structures in the ISM. Secondly, these filamentary structures undergo fragmentation and collapse, leading to the formation of pre-stellar cores, which finally grow into stars. 

Numerous studies have found that magnetic field (B field) may play important roles in the first stage of filament formation within the molecular clouds. Filaments in the Gould Belt revealed with \emph{Herschel} have already been found to be parallel or perpendicular to the B field \citep{Li2013, Palmeirim2013, Cox2016, Panopoulou2017, Carriere2022}. Studies with density structures even suggested that the filaments at low column densities tend to be parallel to the B field, while they become perpendicular at high column densities beyond $A_V$=3 \citep{PlanckCollaboration2016, Jow2018, Fissel2019}. Compared with the numerical simulations \citep{Soler2013, Soler2017}, these suggest that the B field plays a dynamically important role in the formation and evolution of the filaments \citep{Pineda2023}.

The influence of HII regions on filament formation should not be neglected either. \citet{Dobashi1996} found that molecular clouds associated with HII regions are often the birthplace of more massive stars, especially those with mass $\leq 100,M_{\sun}$. \citet{Yamaguchi1999} also reported that the expansion and interaction of HII regions with their surrounding environment play a significant role in their neighboring star formation, a result that is further supported on Galactic scales by the enhanced occurrence of dense clumps and YSOs along the rims of infrared bubbles powered by HII regions \citep{Kendrew2016,Palmeirim2017}. \citet{Inutsuka2015} proposed a scenario of molecular cloud formation driven by expanding bubbles created by supernova remnants or expanding HII regions. In this scenario, molecular clouds are compressed by expanding shock waves, forming filamentary molecular clouds in the shells of the expanding bubbles. Three-dimensional MHD simulations also show remarkable filament-formation signatures on the supershells interacting with HII regions \citep{Dawson2015,Ntormousi2017}. However, to what extent HII regions contribute to the formation of filaments remains poorly understood.

The region we studied, Cygnus X, is one of the most massive giant molecular clouds in our Galaxy, with a mass of about $3\times10^{6}$ $M_{\sun}$ \citep{Cao2019}. It covers a 30-square-degree area in the Galactic plane at a distance of 1.4 kpc \citep{Rygl2010}. Recently, \citet{Cao2019} created a column-density map of the entire Cygnus X region using \emph{Herschel} (70-500 $\mu$m) and unveiled a rich collection of filamentary structures, such as the DR21 filament \citep{Hennemann2012, Ching2022, Cao2022, Li2023}, presenting a valuable opportunity for our filament research. Moreover, Cygnus X is home to numerous distinct HII regions, Wolf-Rayet stars, O-type stars, and OB associations \citep{Cao2019}. Infrared (IR) and millimeter-wavelength large-scale surveys also identified a large number of young stars and protostars \citep{Motte2007, Lucas2008, Beerer2010, Schneider2012, Kryukova2014}. All these make Cygnus X an ideal laboratory for our study of filamentary structures and the correlation between filaments and star formation.

\section{Data analysis and results} \label{sec:Data analysis and results}
\subsection{Filament and core extraction} \label{subsec:filament extraction}

\begin{figure*}[h!]
        \centering
        \includegraphics[width=0.8\textwidth]{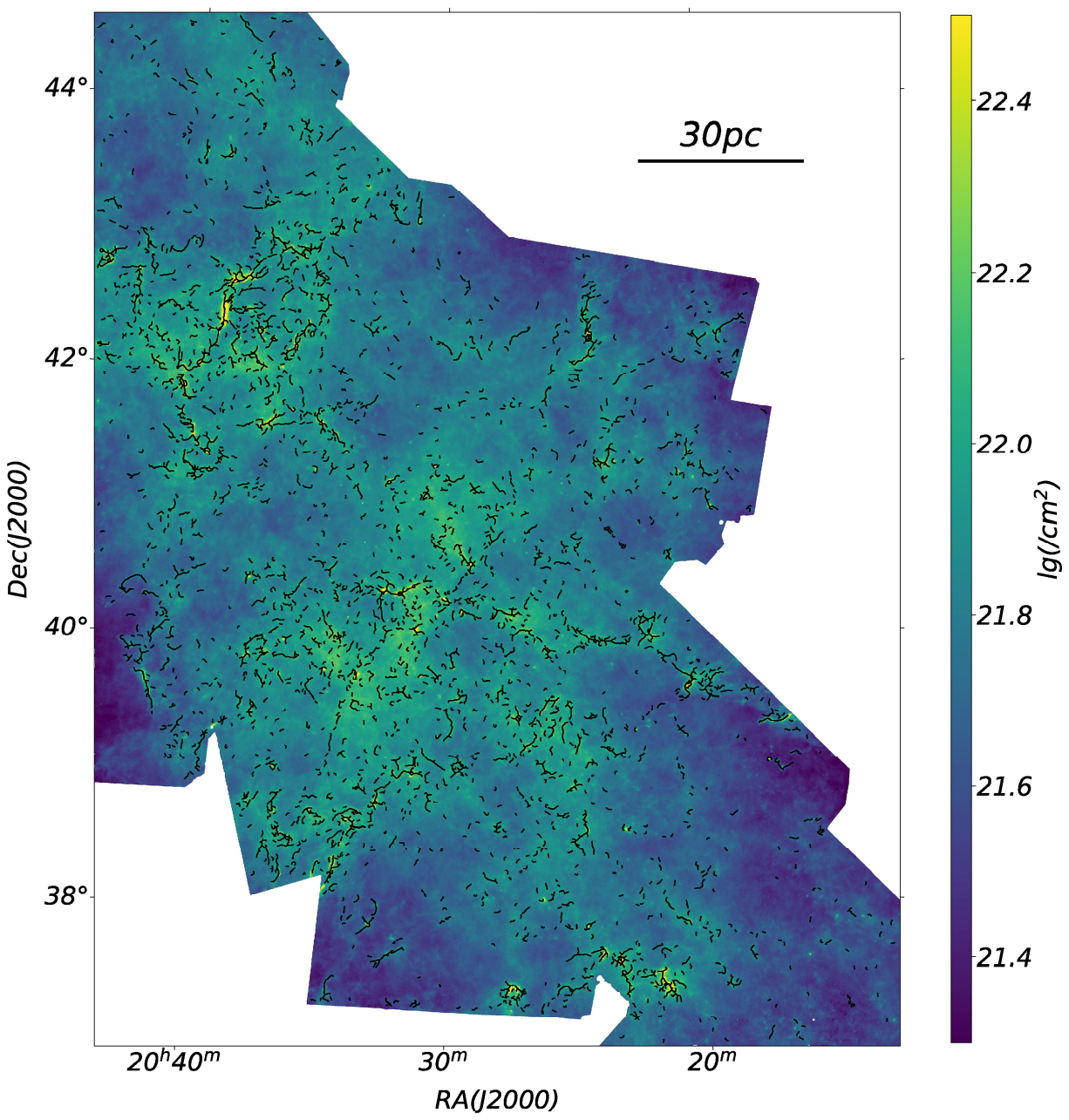}
        \caption{Skeletons of all filaments extracted with \emph{getsf} are shown by black lines. The background is the column-density map of Cygnus X, which was derived by \citet{Cao2019} with the \emph{Herschel} multi-band dust-continuum-emission maps.}
        \label{fig:filament}
    \end{figure*}

We employed the \emph{getsf} algorithm \citep{Menshchikov2021a}, which extracts the sources and filaments by separating their structural components, for the filament and core extraction on the original 18$^{\arcsec}$ column-density map in \citet{Cao2019}. A great advantage of the \emph{getsf} algorithm is that it fully accounts for the nature of multi-scale and varying morphology of molecular clouds, and thus it can decompose the clouds into cores, filaments, and diffuse background.
From the user side, the key parameter to be set is the maximum size of the filaments and cores to be extracted. The filament and core selection criteria in \emph{getsf} are based on the benchmark tests in \citet{Menshchikov2021b}. Here, as a followup work of \citet{Cao2021}, we also focused on cores with a size of approximately 0.1 pc, and thus set the maximum core size to 90" (0.6 pc). For the filament parameter, we performed a round of pre-processing on the data and find a typical width of 0.5 pc for the filaments. Therefore, we set the max size of filament widths to be 126" (0.8 pc), which is sufficient to cover all of the potential filaments. To allow a reasonable measurement of the filament aspect ratio and length, we selected skeletons at least five beam sizes long. We also excluded the filaments and cores within five beams of the map boundaries to avoid the boundary effect. The final filament skeletons are shown in Fig. \ref{fig:filament}. In total, we obtain 6551 cores and 2633 filaments. To double check the feasibility of the filament extraction algorithm, we also extracted the filaments with the other two widely used algorithms, \emph{DisPerSE} and \emph{FilFinder}, in a representative region, and compare the results with that from \emph{getsf}. In general, the three algorithms give consistent results, and more details are presented in Appendix \ref{subsec:Compared with DisPerSE/}. 

\subsection{The orientation, radial column-density profile, and width of the filaments} \label{subsec:filament width}

To quantify the filament orientations, we utilized the principle-component-analysis (PCA) method on seven consecutive pixels (pixel size = 4$^{\arcsec}$) of the skeletons to determine the orientation of the filaments at each skeleton pixel.  

\begin{figure}[h!]
        \centering
        \includegraphics[width=\columnwidth]{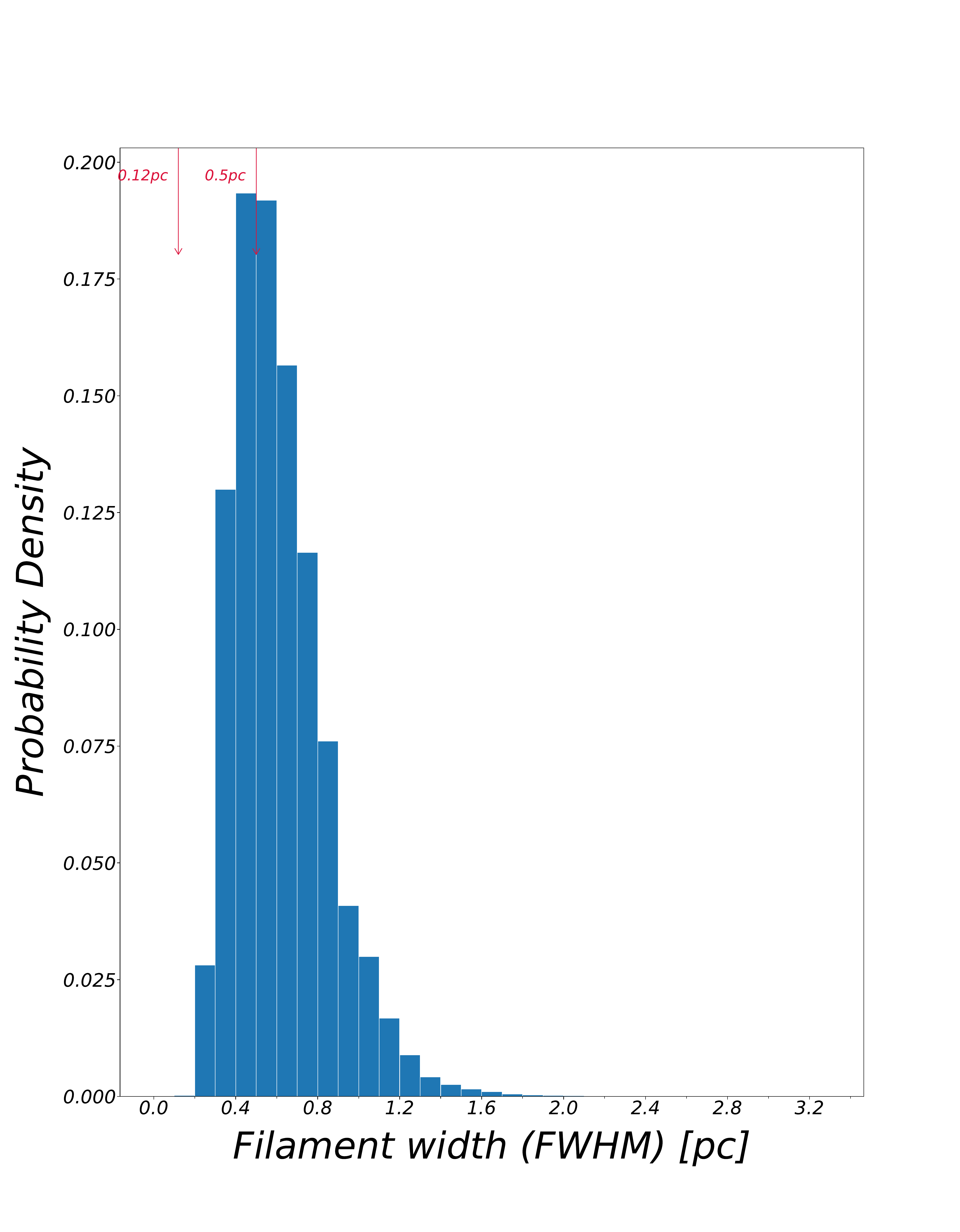}
        \caption{Normalized width distribution of extracted filament skeletons (Fig.~\ref{fig:filament}). This distribution has a peak at 0.5 pc, which indicates a typical width of the filaments in Cygnus X. The resolution 0.12 pc is also labeled as an arrow.}
        \label{fig:width distribution}
    \end{figure}

We then obtained radial column-density profiles along the cuts perpendicular to the filament orientations and fit these profiles with a Gaussian function to measure the full width at half maximum (FWHM) of the filaments. The procedure of the width measurement involves a five-step sequence (see Appendix \ref{subsec:The five-step preprocessing sequence in the width measurement/} for details). Each pixel on the filament skeletons has a width derived from the radial column-density profile, and we obtain a width distribution peaking at $\sim$ 0.5 pc (Fig. \ref{fig:width distribution}). Our column-density map has a pixel size of 4$^{\arcsec}$ and a resolution of 18.4$^{\arcsec}$ \citep{Cao2019}, and thus the histogram shown in Fig. \ref{fig:width distribution} is over-sampled, but the distribution is not affected given the very large sample size. More details about the reliability of filament width measurement are provided in Appendix \ref{subsec:Impact of various models on the fitting qualities and width values/}.

\subsection{Mass functions of cores on and outside of filaments} \label{subsec:filamentary structures and cores and CMF}

\begin{figure}[h!]
        \centering
        \includegraphics[width=\columnwidth]{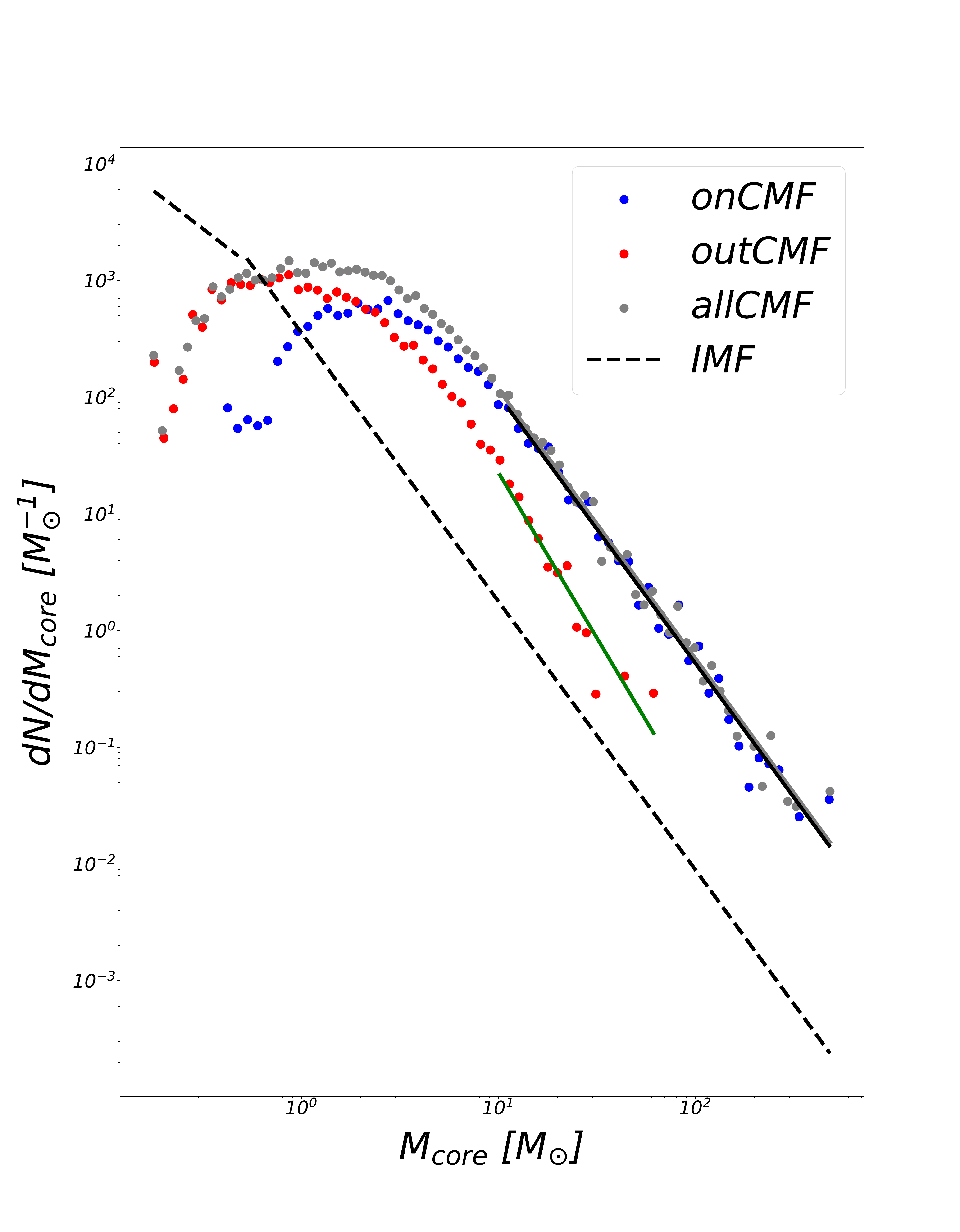}
        \caption{onCMF (blue dots) and outCMF (red dots) derived from the cores on and outside of filaments, respectively (Sect. \ref{subsec:filamentary structures and cores and CMF}). The two different power-law slopes for the two CMFs at the high-mass end (over 10 M$_{\odot}$) are -2.30 and -2.83 for onCMF (black solid line) and outCMF (green solid line), respectively. This analysis highlights a tendency of high-mass cores to be preferentially located on filaments. The allCMF (gray dots) also has the similar power-law slope -2.30 at the high-mass end (over 10 M$_{\sun}$). The initial mass function from \citet{Kroupa2003} is shown by a dashed line.}
        \label{fig:onoutCMF}
    \end{figure}

We checked whether or not the cores are located on the filaments based on this criterion: whether its spatial extent, determined by its size (geometric mean of the major and minor axes of an ellipse), overlaps with the spatial extent of the filament, which is determined by its width. We find that the probability of cores being located on filaments increases with mass, reaching $\textgreater$93$\%$ for those with masses above 20 M$_{\sun}$ (see Tab. \ref{tab:core and filament}). Then, we constructed three CMFs (see Fig. \ref{fig:onoutCMF}): the CMF for cores on filaments (hereafter onCMF), the CMF for cores outside filaments (hereafter outCMF), and the CMF for all of the 6551 cores (hereafter allCMF). The onCMF shows a slope of -2.30, which is derived from the power-law fitting for cores with masses $\textgreater$ 10 $M_\sun$, identical to that of the allCMF. However, the outCMF shows a slope of -2.83 in the same regime (i.e., steeper), which suggests that massive cores are less likely to form outside of filaments.

\begin{table*}[h!]
\caption{\label{tab:core and filament}Cores on filaments}
\centering
\begin{tabular}{cccc}
\hline\hline
Mass&Total Cores (tn)&Cores On The Filaments (cn)&Rate (cn/tn)\\
\hline
$M_\sun$             & Number  & Number   & $100\%$   \\
\hline
\textgreater 20   & 352  & 328   & 93   \\
10-20             & 574  & 476   & 83   \\
5-10              & 1221 & 891   & 73   \\
\textless 5       & 4404 & 2047   & 46   \\
total             & 6551 & 3742   & 57   \\
\hline
\end{tabular}
\tablefoot{The second column is the total number of cores in the certain mass range. The third column is the number of cores on filaments. The fourth column is the probability of a core located on the filaments.}
\end{table*}

\subsection{Filaments and B field} \label{subsec:filament orientation and magnetic field}

\begin{figure*}
\sidecaption
  \includegraphics[width=1.2\columnwidth]{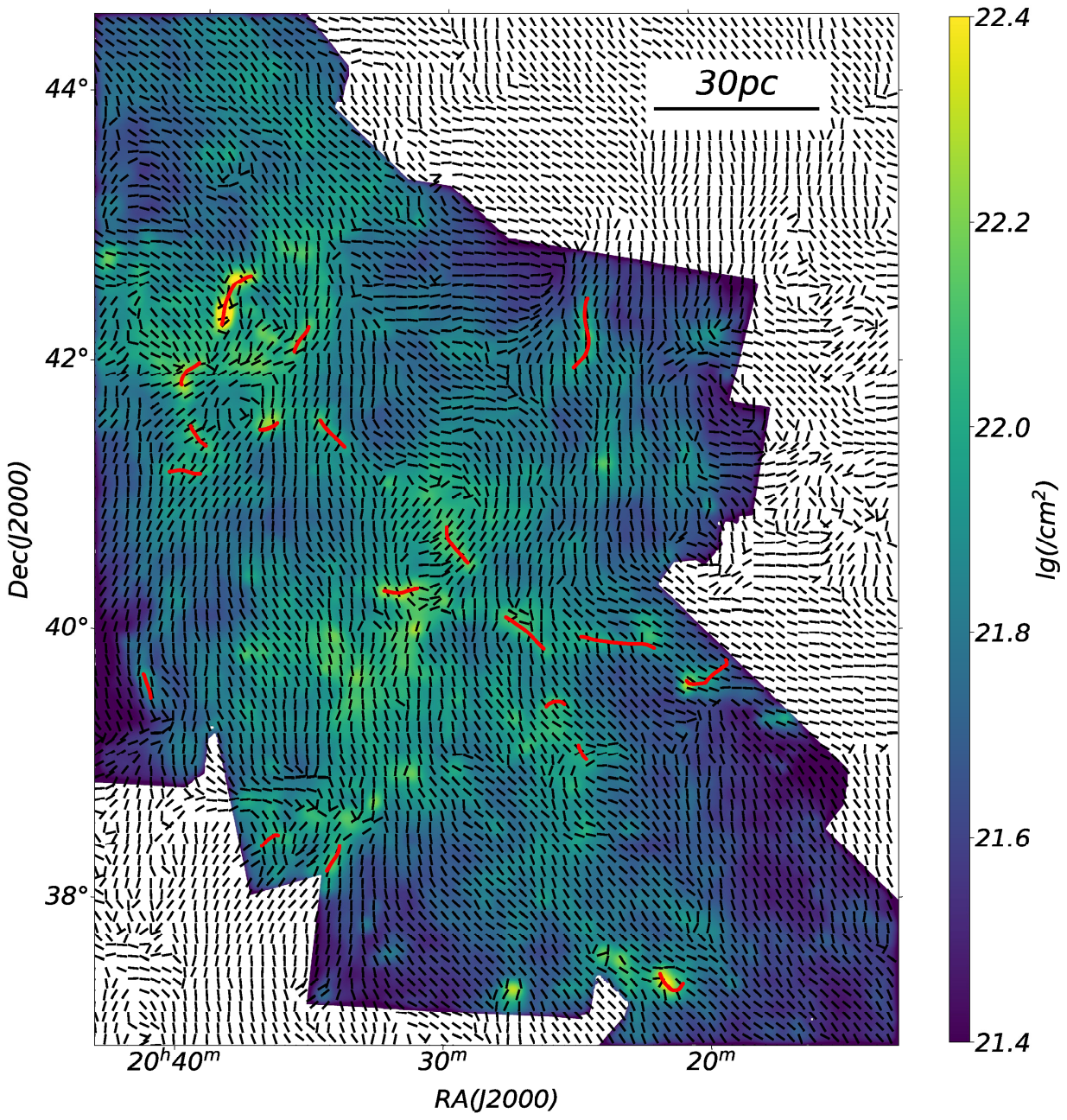}
     \caption{\emph{Planck} B field (black segments), smoothed column-density map (5’,background), and re-extracted filaments (red solid lines) on this smoothed map.}
     \label{fig:planckbone}
\end{figure*}

\begin{figure}[h!]
        \centering
        \includegraphics[width=0.5\textwidth]{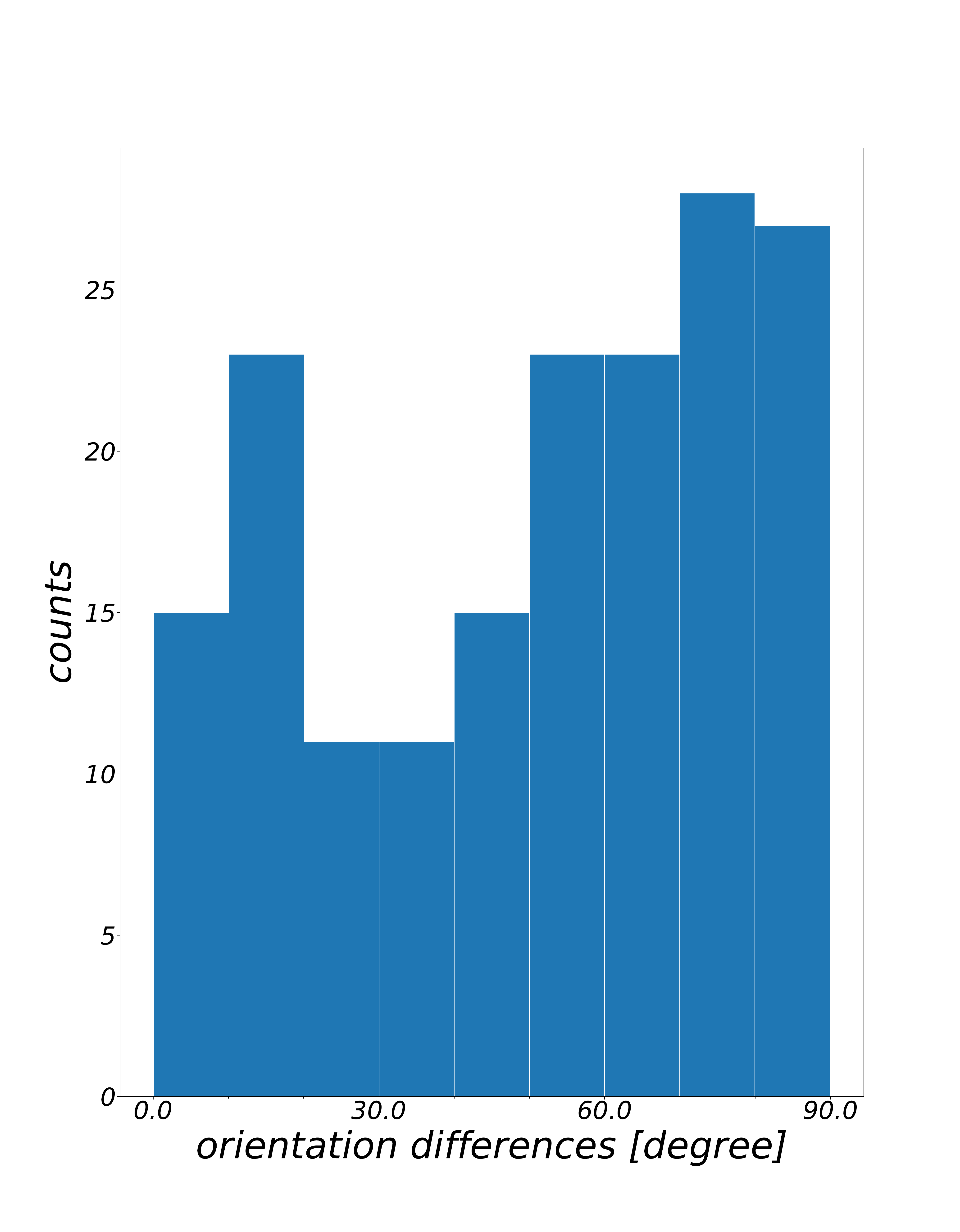}
        \caption{Distribution of orientation differences between filaments and B field at a resolution of 5' (2 pc). This bimodal distribution peaking at $0^{\circ}$ and $90^{\circ}$ suggests filaments tend to be perpendicular or parallel to the B field on a several-pc scale.}
        \label{fig:beam}
    \end{figure}

The orientation of the B field in Cygnus X is derived from the \emph{Planck} 353 GHz dust-polarized emission map (see Fig. \ref{fig:planckbone} and Appendix \ref{subsec:orientation between filaments and magnetic field based on the mismatch resolution/} for more details). The signal-to-noise ratios (S/N) of the \emph{Planck} polarization data is discussed in Appendix \ref{subsec:The signal to noise of the Planck map/}. We smoothed the \emph{Herschel} column-density map to 5' to match the resolution of the \emph{Planck} 353 GHz map and then re-extract the filaments and measure their orientations. Subsequently, we calculated the orientation difference between the newly extracted filaments and the B field per half beam of the \emph{Planck} 353 GHz map. The distribution of the orientation differences is shown in Fig. \ref{fig:beam}, which shows a bimodal distribution peaking at 0$^{\circ}$ and 90$^{\circ}$. Thus the majority of filaments on the $\sim$ 2 pc (5') scale tend to be perpendicular to the B field, and the rest are roughly aligned parallel to the B field. If we compare the filaments derived from the original (18.4$^{\arcsec}$) column-density map with the \emph{Planck} B field map, we also see a preference of the two being perpendicular to each other (Appendix \ref{subsec:orientation between filaments and magnetic field based on the mismatch resolution/}). The smoothed column-density map and the \emph{Planck} B field are also shown in Appendix \ref{subsec:orientation between filaments and magnetic field based on the mismatch resolution/}.

\subsection{Filaments and HII regions} \label{subsec:Filament and HII region}

\begin{figure*}
\sidecaption
  \includegraphics[width=1.2\columnwidth]{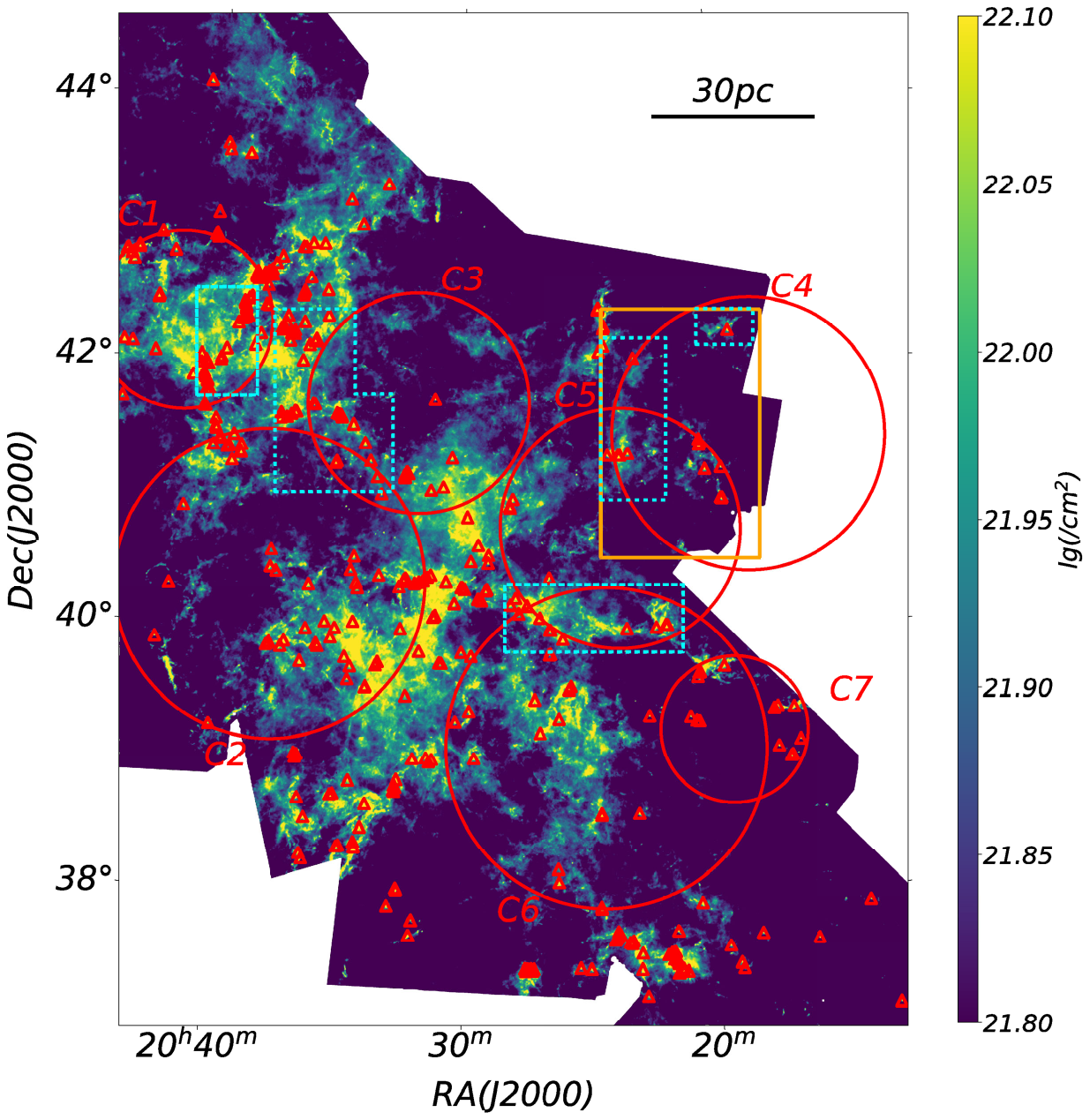}
     \caption{Developed HII regions \citep[red circles numbered 1-7,][]{Cao2019} and massive cores (red triangles) in Cygnus X. The background is the high-contrast column-density map \citep{Cao2019}. The cyan rectangles show four regions where filament ages were computed (Sect. \ref{subsec:Filaments and HII regions/} for details). The orange rectangle to the northwest shows evidence that filaments are compressed by the No.4 HII region (see also Sect. \ref{subsec:Filaments and HII regions/} for details). The majority of high-mass filaments and cores tend to be located on the boundaries of HII regions or the intersections of multiple HII regions, hinting at a strong relationship between the HII regions, filaments, and high-mass cores.}
     \label{fig:HII number}
\end{figure*}

Fig. \ref{fig:HII number} presents an overview of the distribution of developed HII regions \citep{Cao2019}, high-mass cores, and filamentary dense gas structures in Cygnus X. In the high-contrast column-density map, the dense gasses with filamentary morphologies are mostly found in two regions: one surrounded by HII regions C2, C3, C5, and C6; and the other surrounded by C1, C2, and C3. There are clear filamentary gas structures in the intersections between HII regions C5 and C6, and close to the eastern edge of HII region C4. Cores with masses exceeding 20 $M_\sun$ are highly correlated with the high-density filamentary structures, and thus are also found to be mostly located close to the edges of HII regions or at the intersection areas of multiple HII regions. It appears that dense filaments and high-mass cores tend to be located at the boundaries of HII regions or the inter-sections of multiple HII regions.

\subsection{Filaments and young stellar objects} \label{subsec:Filament and YSO}

\begin{figure*}[h!]
    \centering
    \includegraphics[width=0.8\textwidth]{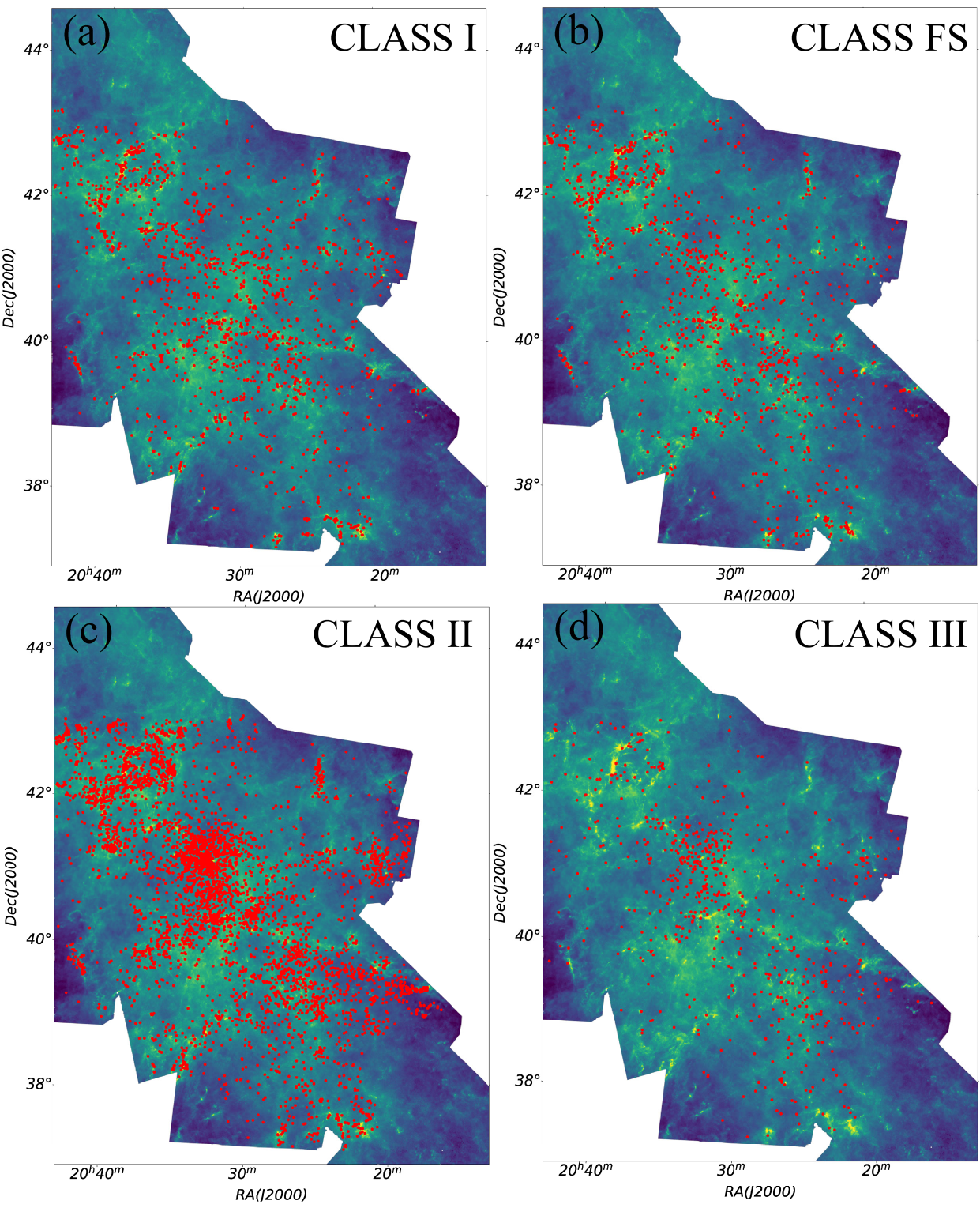}
    \caption{Spatial distributions of YSOs (red dots) in four evolution stages: a) CLASS I; b) CLASS FS; c) CLASS II; d) CLASS III. The background is the \emph{Herschel} column-density map. YSO data were obtained from \citet{Kuhn2021}. As YSOs evolve, they tend to gradually move away from filaments.}
    \label{fig:YSOCLASSDISTRIBUTION}
    \end{figure*}

\begin{figure}[h!]
        \centering
        \includegraphics[width=\columnwidth]{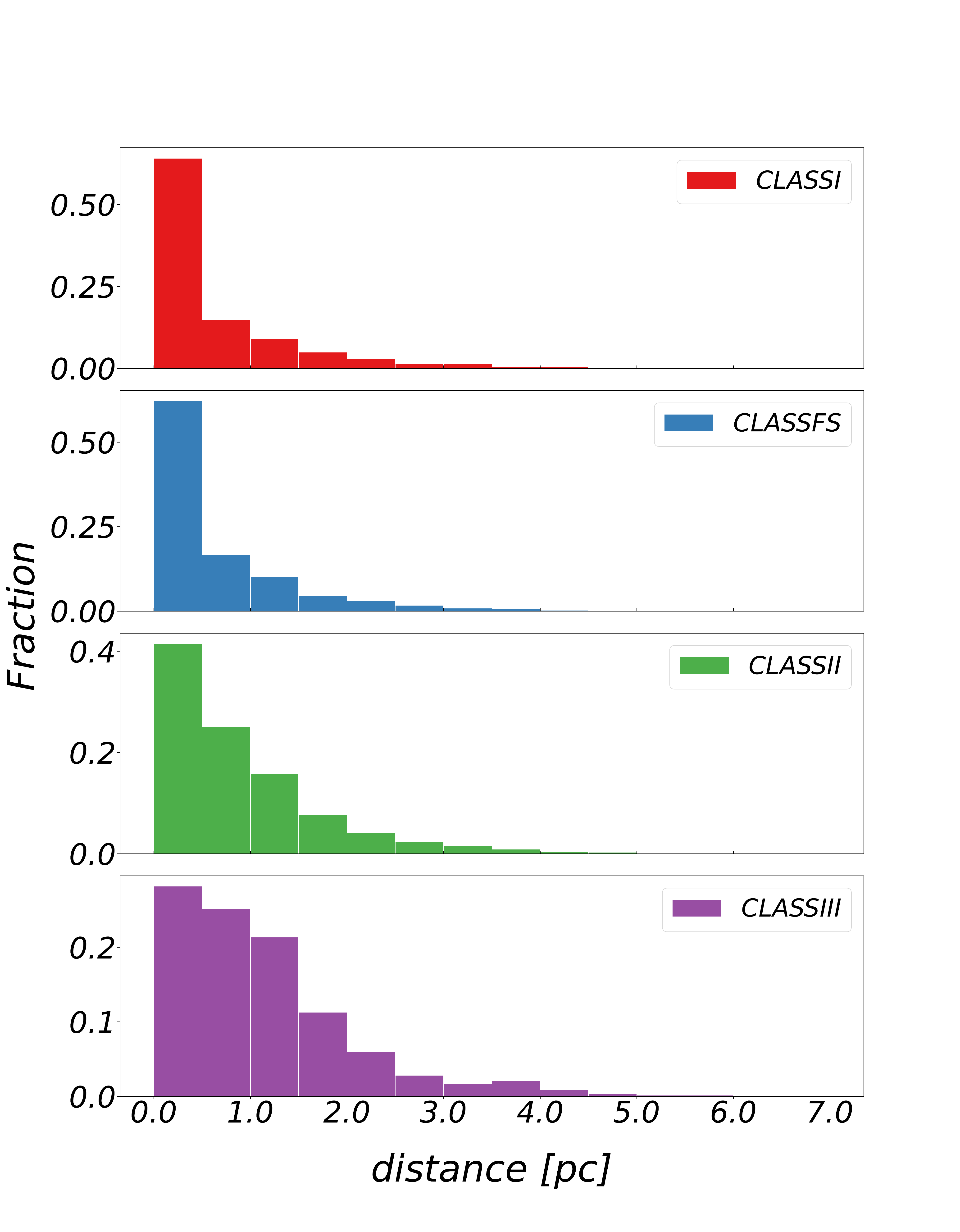}
        \caption{Distributions of distances between YSOs and their nearest filaments; also shown as the percentage of sources at each distance for different evolutionary types. From top to bottom, the panels show CLASS I, CLASS FT, CLASS II, and CLASS III, respectively. The bin size is 0.5 pc, which is the typical width of the filaments in Cygnus X. There is a clear trend of YSOs gradually moving away from the nearest filament with the evolution.}
        \label{fig:classdistance}
    \end{figure}

Given that the filaments are ubiquitously found in Cygnus X, it is of great interest to check whether the spatial distribution of young stellar objects (YSOs) are correlated with the filaments. We overlay identified YSOs at various evolutionary stages \citep{Kuhn2021} on our column-density map (Fig. \ref{fig:YSOCLASSDISTRIBUTION}). We determine a YSO to be on a filament when the YSO is located within the FWHM of the radial column-density profile of the filament. We calculated the rates of being on the filaments for the YSOs at each evolutionary stage (Tab. \ref{tab:YSOs and filaments}). It is immediately clear that younger YSOs have a greater probability of being on a filament. From Fig. \ref{fig:YSOCLASSDISTRIBUTION}, we can see that Class I and Class flat-spectrum (FS; YSOs, whose infrared spectral index lies between those of Class I and Class II objects) YSOs are to some extent spatially correlated with the filamentary gas structures, and that correlation is less evident for Class II YSOs and not seen at all for Class III YSOs. We also calculated the distance between a YSO and its nearest filament skeleton. Considering that the extracted filaments here have a typical FWHM width of 0.5 pc, it again shows that younger YSOs are more likely to be on the filaments (Fig. \ref{fig:classdistance}).

\begin{table*}[h!]
\caption{\label{tab:YSOs and filaments}YSOs and filaments}
\centering
\begin{tabular}{cccc}
\hline\hline
Stage&Total YSOs (tn)&YSOs On The Filament (cn)&Rate (cn/tn)\\
\hline
            & number  & number   & $100\%$   \\
\hline
CLASS I      & 2097  & 1120   & 53   \\
CLASS FS     & 1532  & 768   & 50   \\
CLASS II     & 4829  & 1217   & 25   \\
CLASS III    & 674 & 86   & 13   \\
\hline
\end{tabular}
\tablefoot{The second column lists the total number of YSOs in that stage. The third column is the number of the YSOs on the filaments. The fourth column is the probability of a YSO located on the filaments.}
\end{table*}

\section{Discussion} \label{sec:Discussion}

\subsection{The typical width of filaments} \label{subsec:Filaments and Width/}

The filaments extracted from the column-density map of Cygnus X show a typical width of 0.5 pc with a resolution of $\sim$ 0.12pc. Compared to the HGBS surveys, the filaments are measured at a lower linear resolution, but the derived typical width is about four times the resolution, and thus it cannot be solely due to the resolution-limitation effect. This value is different from the width of 0.1 pc for the filaments seen in the Gould Belt survey \citep{Arzoumanian2011,Arzoumanian2019}. Other observations also reported different filament widths measured with various resolutions, such as G035.39-00.33 \citep[$\sim$0.028 pc;][]{Henshaw2017},
L1287 \citep[$\sim$0.03 pc;][]{Sepulveda2020},
Orion \citep[$\sim$0.035 pc for ISF;][]{Hacar2018},
Aquila$\&$Polaris \citep[$\sim$0.04 pc;][]{Menhchikov2010},
G14.225-0.506 \citep[$\sim$0.05–0.09 pc;][]{Chen2019},
Musca \citep[$\sim$0.07 pc;][]{Kainulainen2016}, DR21 \citep[$\sim0.26-0.34$ pc;][]{Hennemann2012}, Taurus \citep[$\sim$0.4 pc;][]{Panopoulou2014}, and Galactic-Plane filaments \citep[$\sim$0.1-2.5 pc;][]{Schisano2014}. \citet{Hacar2023} also indicated that distance determination, physical processes, tracers and biases in the measurement method all influence the measurement of filament widths. These results strongly suggest that molecular gas filaments have a wide range of sizes and densities, and they can be hierarchical structures. In our study of Cygnus X, we thus most likely picked up molecular filaments with a typical width of 0.5 pc and a length of a few pc with \emph{getsf}. In terms of scale, these filaments seem to bridge their parent molecular clouds \citep[with a scale of tens pc and an average density 10$^{2}$ cm$^{-3}$ - 10$^{3}$ cm$^{-3}$, also see][]{MivilleDeschenes2017} and dense cores \citep[with a scale of $\sim$0.1 pc and an average density $>$10$^{4}$ cm$^{-3}$,][]{Cao2021, Bergin2007} in Cygnus X.  

\subsection{Evidence for massive core formation preferentially on filaments} \label{subsec:Filaments and CMF/}

Cao et al. (2021) extracted all the dense cores in Cygnus X using \emph{getsources}, which is a predecessor of \emph{getsf}, and yielded a power-law distribution with a slope of -2.30 for the cores of masses $>10~M_{\odot}$. Here, with an updated sample of cores obtained with \emph{getsf}, we find that the high-mass ($>10~M_{\odot}$) regime of allCMF is -2.30, confirming the results of \citet{Cao2021}. More interestingly, in the same mass range, the onCMF has a power-law slope of -2.30 as well, indicating that the mass function of high-mass cores is overwhelmingly determined by the cores on the filaments. The outCMF, on the other hand, has a much steeper power-law distribution for the high-mass cores, meaning a sharp decrease in the possibility of finding a core outside of filaments as the core mass increases. This trend is further quantified in Tab. 1. As the core mass goes up to $>20~M_{\odot}$, 93$\%$ of cores are located on filaments, and for cores with masses of $>70~M_{\odot}$, they can only be found on filaments (Fig. \ref{fig:onoutCMF}). It implies that filaments are fertile ground for the formation of more massive cores, and they may provide a mass reservoir for accretion onto high-mass cores. Such a scenario is consistent with \citet{Li2021}, which found that massive cores in Cygnus X grow mostly through accretion.

\begin{figure}[h!]
        \centering
        \includegraphics[width=0.5\textwidth]{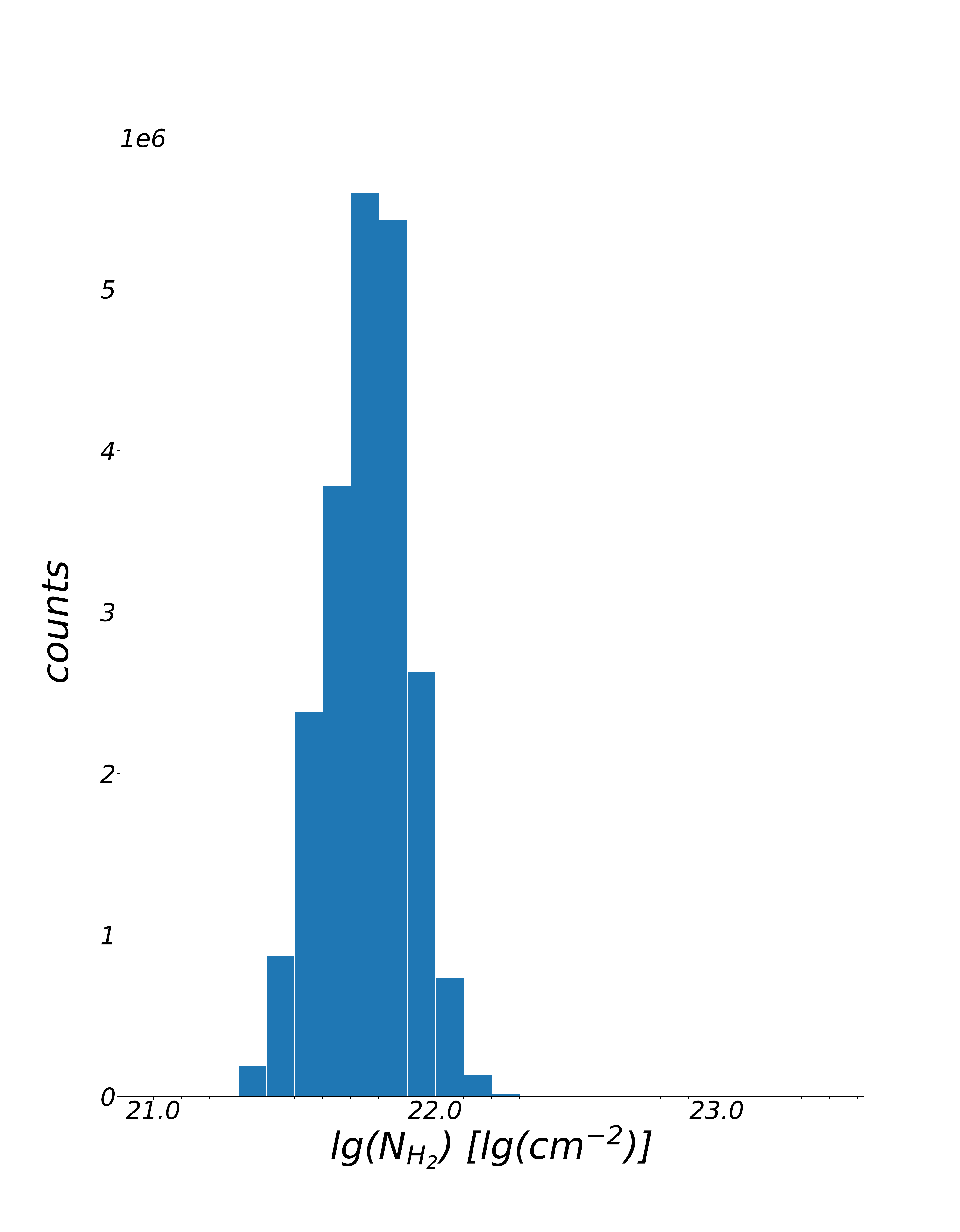}
        \caption{The column-density distribution for regions outside the filaments. This distribution is narrowly peaking at 5 $\times10^{21}$ cm$^{-2}$, which indicates a typical column-density for regions outside the filaments.}
        \label{fig:outfilamentdistribution}
    \end{figure}

\citet{Andre2014} found that filaments ubiquitously exist in the Aquila molecular cloud complex and that the Bonnor-Ebert mass \citep{Bonnor1956} is very close to the peak of the pre-stellar CMF in Aquila, implying a physical connection between filament fragmentation and the peak of the CMF. In Cygnus X, the peak of onCMF differs from that of outCMF, and it is evident that the environments in which onCMF and outCMF arise have significantly different densities; thus, we estimated the local Bonner-Ebert masses on and outside of filaments to explore whether the difference between the peaks of the two CMFs could be fully explained by gravitational fragmentation under different physical conditions. The local critical Bonnor-Ebert mass\citep{Bonnor1956} can be derived as:
\begin{equation}
M_{\rm BE,th}\sim0.5M_{\odot}(\frac{T}{10K})^{2}(\frac{\sum_{\rm cl}}{160M_{\odot}pc^{-2}})^{-1},
\end{equation}
where $\sum_{\rm cl}$ is the surface density and T $\sim$ 10 K is the typical gas temperature in Cygnus X. Fig. \ref{fig:outfilamentdistribution} shows the column-density distribution for regions outside the filaments, which reveals a peak value of $\sum_{\rm cl}$ $\sim$ 5 $\times$ $10^{21}$ cm$^{-2}$. We treat this value as a typical column density for regions outside the filaments and thus derive a Bonnor-Ebert mass of $M_{\rm BE,th,out}\sim0.8$ $M_{\odot}$, which is close to the peak of outCMF ($\sim0.5$ $M_{\odot}$). For onCMF, $\sum_{\rm cl}$ must be larger than that of outCMF. According to the equation (2), $M_{\rm BE,th}$ is $\propto$ $\sum_{\rm cl}^{-1}$, and thus the Bonnor-Ebert mass $M_{\rm BE,th,on}$ must be less than $M_{\rm BE,th,out}$, and certainly much less than the peak of onCMF ($\sim$ 2 $M_{\odot}$). This difference between $M_{\rm BE,th,on}$ and the peak of onCMF indicates that cores on the filaments could have accreted a significant fraction of material again, consistent with the finding of \citet{Li2021}. In Orion A, \citet{Polychroni2013} similarly reported a peak mass of 4 $M_{\sun}$ for their onCMF and a peak of 0.8 $M_{\sun}$ for their outCMF.

\emph{Herschel} studies of nearby clouds have already shown a strong spatial correlation between filaments and pre-stellar cores: approximately 75$\%$ of pre-stellar cores are located in dense filaments \citep{Andre2010, Konyves2015, Andre2017}. Subsequent studies also indicate that over 80$\%$ of candidate pre-stellar cores are located on the crests of the nearest filaments \citep{Konyves2015, Andre2017, Bresnahan2018}. Through this work, with a total number of 6551 cores and 2633 filaments in the same region, we see that 57$\%$ of total cores are located on filaments, and the rate continues to increase with the core mass. Our findings emphasize that the core mass growth (and thus the formation of massive cores) is strongly dependent on the presence of filaments. Future studies that could determine the nature of the cores, i.e., starless, pre-stellar, and proto-stellar, would provide deep insights into the question of how the filaments induce star formation.
 
In addition, the mass function of molecular clouds (MCs) and CO clumps within MCs is $\Delta N / \Delta M \propto M^{-1.6 \pm 0.2}$, with a shallow profile \citep{Solomon1987, Blitz1993, Kramer1998, Rice2016}, and significantly shallower than the Salpeter IMF with a power-law index of -2.35 \citep{Salpeter1955}. The filament mass function (FMF) and filament line mass function (FLMF) has also been constructed in a comprehensive study of filament properties with HGBS observations \citep[][see their Fig. 1]{Andre2019}. Both the FMF and FLMF have an index of $\sim$ $-2.4$ at the high-mass (high-mass line) end, which is similar to that of CMFs. Overall, the cloud mass function is significantly shallower than both the FMF (FLMF) and CMF. However, at the high-mass end the FMF (FLMF) and CMF share a common slope of approximately –2.4. This similarity implies that the CMF’s high-mass tail is inherited from the filamentary mass distribution, indicating that the formation of the most massive cores is tightly regulated by their parent filaments.

\subsection{Alignment between filaments and B field} \label{subsec:Filaments and B field/}

\begin{figure}[h!]
        \centering
        \includegraphics[width=0.5\textwidth]{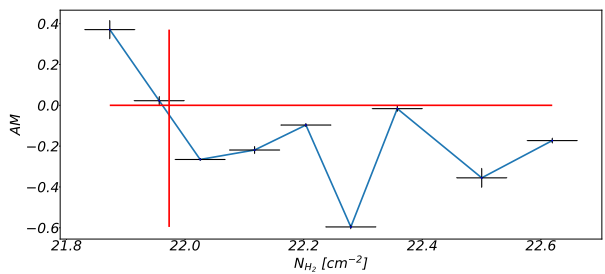}
        \caption{AM versus column-density. The horizontal line is the zero line and marks the boundary between the parallel case (AM $\textgreater$ 0) and the perpendicular case (AM $\textless$ 0). The vertical line marks the position of TN, which is around $A_V$=10. The uncertainties of the AM and the column-density for each data point are shown as red error bars (see Appendix \ref{subsec:The relationship between the orientation differences and the densities/} for further details). The trend is clear here: filaments and the B field tend to be parallel below TN, while they become perpendicular beyond TN.}
        \label{fig:bin10}
    \end{figure}

The relative orientations between the filaments and B field on the $\sim$2 pc scale show a bimodal distribution (Fig. \ref{fig:beam}). Similar results are found for filaments in nearby molecular clouds \citep{Li2013, Carriere2022}. We further explore how this distribution is related to the column density. We define a parameter, the alignment measurement (AM) as <2cos$^{2}\theta$-1>, where $\theta$ is the absolute difference between the filament and B-field orientations (\citealp{Lazarian2018} and see Appendix~\ref{subsec:The relationship between the orientation differences and the densities/} for more details). When AM $\textgreater$ 0, the filaments are oriented more parallel to the B-field orientation, and when AM $\textless$ 0, the filaments are more likely to be perpendicular to the B field. Fig. \ref{fig:bin10} shows a trend from parallel to perpendicular with the increase of the filament column densities. Previous observational studies of relative orientation between column-density structures and B field have also reported a similar trend changing from mostly parallel at low densities to preferentially perpendicular at higher densities \citep{PlanckCollaboration2016, Jow2018}. In Fig. \ref{fig:bin10}, the transition from parallel to perpendicular occurs at a column density of $N_{\rm {H_2}}\sim0.94\times10^{22}$ cm$^{-2}$, corresponding to a visual extinction of $A_V=10$ mag for $N_{\rm H}/A_V=1.87\times10^{21}$ cm$^{-2}$ \citep{Bohlin1978, Vrba1984}. This transition column-density, TN, is very close to a column-density threshold of $A_V=8$ mag, above which pre-stellar cores effectively form and star formation could happen, as found in nearby molecular clouds \citep{Lada2010, Andre2014}. Thus, the TN here indicates that the dense and star-forming filaments are perpendicular to the B fields. Note that \citet{PlanckCollaboration2016} reported a TN of $A_V=3$, but there the density structures are derived from the local column-density gradient $\nabla{\Sigma}$. In a simplified case of an isolated filament with a radial column-density profile peaking at the filament spine, $\nabla{\Sigma}$ is expected to be orthogonal to the filament orientation, but in more general cases with complicated column-density profiles and networked filaments, the filament orientation could not be by probed by $\nabla{\Sigma}$ (see Appendix \ref{subsec:The difference between the filament and the density structure/}). Thus, the TN seen in Fig. 9 is not to be directly compared to that found by \citet{PlanckCollaboration2016}. On the other hand, \citet{Fissel2019} compared the B-field orientation to the orientation of elongated gas structures in Vela C inferred from the zeroth-moment maps of several molecular spectral lines probing a range of gas densities, and they found that low-density structures are more likely to be parallel to the B fields, while high-density structures show a tendency to be perpendicular to the B fields. 

We also calculated the orientation differences between the \emph{Planck} B field and the filaments extracted from the original column-density map (see Appendix \ref{subsec:orientation between filaments and magnetic field based on the mismatch resolution/}). The results show a weak tendency of the filaments to be perpendicular to the B field. Considering the large difference between the resolutions of the B field map (5') and the column-density map (18.4$^{\arcsec}$), the tendency should be taken with caution. However, it is clear from Fig. \ref{fig:beam} that if we compare the filaments extracted from the smoothed column-density map with a resolution identical to that of the B field map, the filaments are predominantly perpendicular to the B field. Taken together, these results suggest that in the Cygnus X region, high-density filaments are generally oriented perpendicular to the magnetic field. The B field of the parent cloud may play a significant role in regulating the formation of filaments within molecular clouds.

\citet{Abe2021} classified theoretical models of filament formation into five types, of which the type-G, -C, and -O modes have filaments perpendicular to the B field \citep[also see][]{Pineda2023}. Type-G is a fragmentation process of a shock-compressed sheet-like cloud induced by gravitational instability; type-C has an initial gas density and B field setup similar to that of type-G, but the filament formation is induced by turbulence, causing converging gas flows along the field lines; in the type-O model, filaments form at the tip of converging flows along an oblique MHD shock front that is created by a shock wave sweeping a highly clumpy cloud. According to the simulation by \citet{Abe2021}, when the velocity of a shock compressing the molecular cloud exceeds V$_{\rm cr}$ $\sim$ 5 km s$^{-1}$ (depending on the density and magnetization level), the type-O mode dominates the formation of filaments within the shock-compressed layers. This model tends to produce more massive and prominent filaments, with column densities exceeding 10$^{22}$ cm$^{-2}$, compared to type-G or -C modes (see their Figs. 2, 3, 6, and 7). In Cygnus X, the expansion of developed HII regions exerts strong influence on the surrounding medium. At the boundaries of these expanding bubbles, numerous dense and prominent filaments with column densities above 10$^{22}$ cm$^{-2}$ are observed. In Figure \ref{fig:HII number}, the molecular cloud complex shows a clearly inhomogeneous distribution of gas density, which is a natural consequence of past and ongoing star formation. Such inhomogeneity is favored by the type-O mode. It is therefore reasonable to speculate that the expansion of large bubbles induces shocks compressing magnetized molecular gas into layers and then forming pc-scale filaments at the tips of the converging flows in the type-O mode. Thus, we suggest that the type-O mode is the most plausible mechanism for filament formation in Cygnus X. In addition, type-G filaments generally have widths smaller than their corresponding Jeans length \citep{Nagai1998}, but the observed filament widths in Cygnus X are much larger, typically 0.5 pc, than the estimated Jeans length ($\sim$ 0.03 pc for column densities above 10$^{22}$ cm$^{-2}$ and a temperature of $\sim$ 10 K). This suggests that the type-G mode may not be the dominant mechanism of filament formation in Cygnus X. For the type-C mode, the turbulent velocity perturbation should be sub-Alfvénic, or at most trans-Alfvénic \citep{Padoan1999, Chen2014}, and thus future observations capable of revealing the velocity structure of the whole complex may provide further insights into this mechanism.   

\subsection{HII regions and filaments} \label{subsec:Filaments and HII regions/}

Various physical processes can compress the cloud gas on a pc scale, as large-scale flows, stellar-wind bubbles, expanding HII regions, or supernova remnants \citep{Elmegreen1998, VazquezSemadeni2007, Whitworth2007}. \citet{deGeus1992} presented a model to interpret the formation of the molecular clouds triggered by the compression of wind-blown bubbles. \citet{Tachihara2001} and \citet{Schneider2006} adopted this model to interpret the formation of the Ophiuchus complex and the complexes around Cyg OB2. \citet{Li2023} also reported a bow-like atomic HI filament harboring a dense molecular filament in the DR 21 region and found that the filament forms at intersection of two expanding HII regions, G081.920+00.138 and Cygnus OB2. Evidence of star formation triggered by HII regions has also been reported \citep{Zavagno2007, Zavagno2010, Deharveng2010}.

Early in the statistical study by \citet{Kendrew2012}, a significant overdensity of massive YSOs near the edge of bubbles was found in the Red MSX Source survey. \citet{Palmeirim2017} found the clump-formation efficiency of the Hi-GAL clumps to be approximately two times higher for those located inside the bubbles than those located outside the bubbles. The observations of Cygnus X show a clear spatial correlation between dense filaments, high-mass cores, and HII regions (Sect. \ref{subsec:Filaments and HII regions/}). All these suggest that expanding HII regions may have a significant impact on the formation of dense gas structures and stars in the surroundings.

\begin{figure*}[h!]
    \centering
    \includegraphics[width=0.8\textwidth]{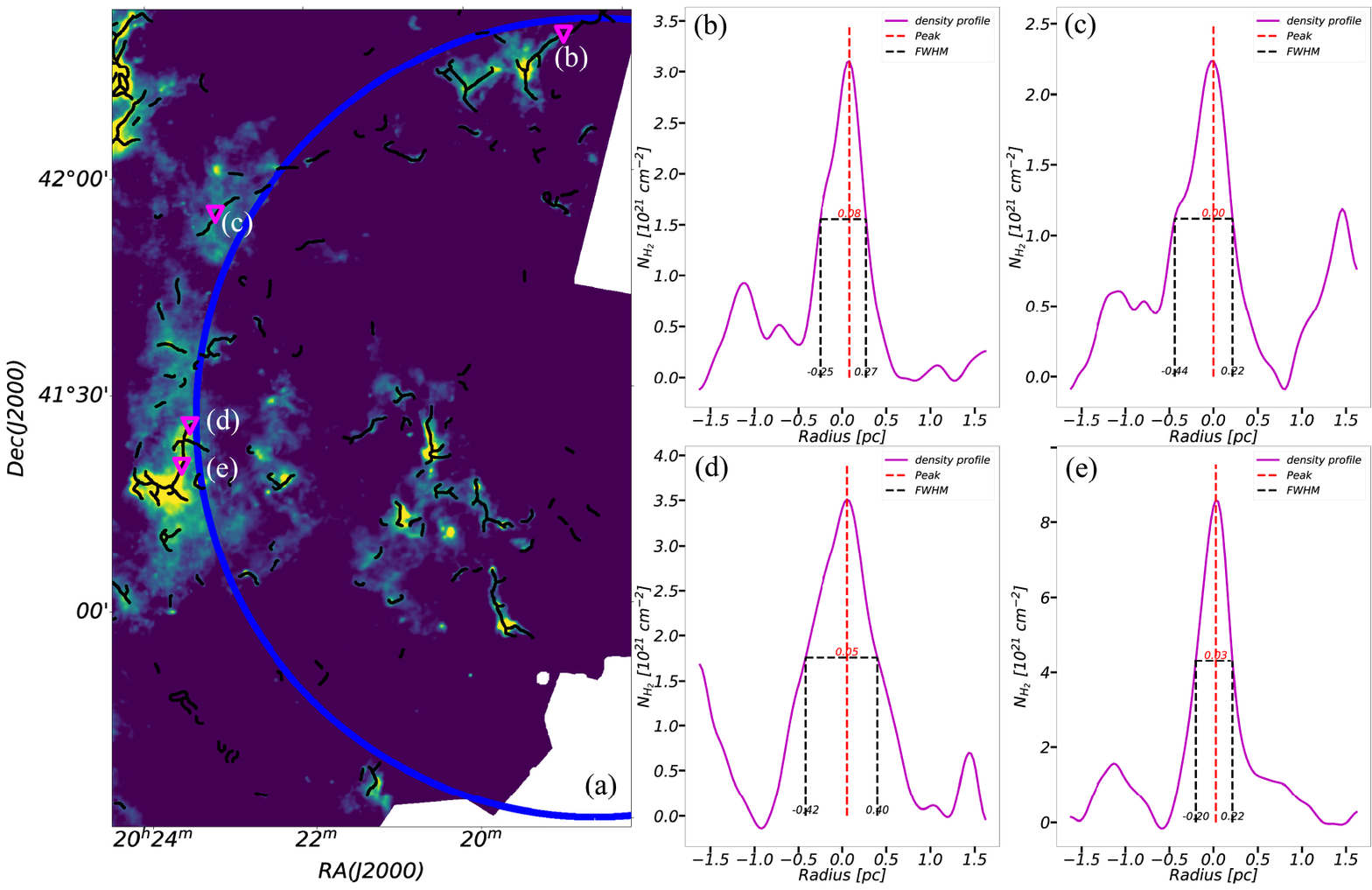}
    \caption{a) No.4 HII region (blue circle) and the surrounding filaments (black solid lines, see Fig. \ref{fig:HII number} for details). The four inverted magenta triangles symbolize the positions where the column-density profiles were extracted; and they are shown in the right panels. b) to e) Four column-density profiles at corresponding positions as labeled in the left panel. The vertical dashed red line marks the position of the peak column-density, while vertical dashed black lines indicate the radii at which the column-density drops to half of its maximum value. The numerical values annotated along the vertical lines specify the corresponding radii of the two half-maximum points, as well as the peak position. Note that the column-density profiles show a steeper gradient toward the HII region (right side of the peak) and a flatter gradient away from it (left side of the peak), suggesting that the filaments are compressed by the expanding HII region. The widths measured at the half maximum on the left sides of the peaks are 0.33 pc, 0.44 pc, 0.47 pc, and 0.23 pc greater than 0.19 pc, 0.22 pc, 0.35 pc, and 0.19 pc on the right for panels b) to e), respectively.}
    \label{fig:HII density profile}
    \end{figure*}

Further evidence of the impact of HII region expansion on the filament formation in Cygnus X is found by checking the filament column-density profiles (Fig. \ref{fig:HII density profile}). The filament profiles become steeper on the side facing the No. 4 HII region (i.e., to the right of the peak) and remain flatter on the opposite side (to the left of the peak). This asymmetry clearly indicates that the filaments are being compressed by the central No. 4 HII region. Similar asymmetric column-density profiles are also seen in RCW120 \citep{Figueira2017, Zavagno2020}. It is therefore reasonable to speculate that the filaments may form from the gas compression by HII region expansion and this compression can also facilitate the growth of high-mass cores. 

To further test the scenario of filament formation by HII region expansion, we try to compare the ages of HII regions and filaments. The calculation of HII region ages is based on the Hosokawa-Inutsuka model \citep{Bisbas2015}: 
\begin{equation}
R_{\rm HII}(t)=R_{\rm St}(1+\frac{7}{4}\sqrt{\frac{4}{3}}\frac{c_{\rm i}t}{R_{\rm St}})^{4/7},
\end{equation}
where $R_{\rm HII}$ is the radius of a developed HII region,  $c_{\rm i}$ is the sound speed in the ionized HII region, and $R_{\rm St}$ is the initial Str\"omgren radius, which is associated with the central OB star UV photon radiation capability and the volume density of the neutral gas around the OB star. However, it is difficult to determine the exact number and spectral types of the OB stars exciting each HII region. Therefore, 1) we estimated the ages of all HII regions (1–6) surrounding the filament enclosed by the cyan rectangles, and 2) we assumed that each HII region is excited by a single OB star. Using the initial Strömgren radii corresponding to all O-type and early B-type stars with luminosity classes V, III, and Ia from \citep{Vacca1996}, we treated this set of values as a plausible range of initial Strömgren radii and substituted it into our calculations. With these assumptions, we derived a rough age range of approximately 1–100 Myr for HII regions 1–6 in Fig. \ref{fig:HII number}. Alternatively, we also calculated the ages of HII regions through another simple consideration---the time required for sound waves to traverse the ionized region. This estimation yields the ages of HII regions $\sim$ $r_{\rm HII}/c_{\rm i}\sim$ 1-2 Myr, where we find $c_{\rm i}$ $\sim10$ km s$^{-1}$ in HII regions. In addition, based on detailed spectroscopic modeling of O stars and HRD isochrone fitting with Gaia DR2, \citet{Berlanas2020} concluded that Cygnus OB2 experienced multiple star-formation bursts, with star formation occurring over an approximate timescale of 1–6 Myr. This timescale is consistent with the ages we derive for the H II regions. Therefore, we consider that our simple age estimates are reasonable.

The ages of the filaments were approximated with the ages of the YSOs in the filaments, and the latter is calculated from the disk fraction \citep{Haisch2001}. The YSO disk fraction is defined as n(Class I + Class II + Class FT)/n(Class I + Class II + Class III + Class FT), where n is the number of YSOs in these stages. The calculated ages of the filaments on the boundaries of the HII regions are $\textless$ $3\times10^{5}$ yr (Fig. \ref{fig:HII number}), younger than the ages of the HII regions (regardless of the method of calculation). This indicates that it is possible that the filaments form through the gas compression by the expanding HII regions.

The overall picture of HII bubbles and filaments in Cygnus X appears to be consistent with the bubble-filament paradigm originally proposed for molecular cloud formation \citep{Inutsuka2015, Pineda2023}. In this framework, expanding HII regions compress the surrounding gas and give rise to high-density filaments along their boundaries. The formation of these filaments presumably happens via the type-O mode, although contributions from type-C or type-G mechanisms cannot be fully ruled out (see Sect. \ref{subsec:Filaments and B field/} of this paper and Fig. 20 in \citealp{Pineda2023} for a schematic illustration).

\section{Summary} \label{sec:Summary}

Using the \emph{Herschel} column-density map, we conducted a complete survey of the filaments in Cygnus X. We further investigated the physical properties of the filaments and their relationships with the cores, B field, YSOs, and HII regions. Our results and conclusions are summarized as follows.

\begin{enumerate}
   \item We identified filaments using \emph{getsf} on a $5^{\circ}\times6^{\circ}$ \emph{Herschel} column-density map. This provided an unprecedentedly large sample of 2633 filaments in Cygnus X.
   \item We derived a filament width distribution peaking at $\sim$ 0.5 pc. Together with other different width measurements, these suggest that filaments are hierarchical structures with various sizes, widths, and mean densities.
   \item We found that the probability of cores being located on filaments increases with mass, reaching $\textgreater$93$\%$ for those with masses above 20 M$_{\sun}$. We further constructed three core mass functions, onCMF, outCMF, and allCMF, based on the spatial relationship between filaments and molecular cloud cores. In the high-mass regime ($\textgreater$ 10 $M_{\odot}$), both onCMF and allCMF exhibit a power-law slope very close to that of the IMF ($-2.30$), whereas outCMF has a much steeper slope ($-2.83$). In addition, by comparing the core mass with the Bonnor-Ebert mass, we found that the cores outside the filaments may form from cloud fragmentation, whereas the cores on the filaments could have accumulated a large fraction of their mass via accretion from the filaments. All this suggest that filaments are the primary sites for forming more massive cores and play a crucial role in forming high-mass cores in Cygnus X.
   \item On the smoothed column-density map (5'), the extracted filaments are preferentially oriented perpendicular to the B field, except that those of lowest column densities are parallel to the B field. Even the filaments extracted from the original column-density map show a weak preference for being perpendicular to the B field, despite the large difference between the resolutions of the column-density map and \emph{Planck} B-field map. When considering the column densities of the extracted filaments and the impact of HII regions in the complex, we suggest that the type-O mode, filaments form at the tip of converging flows along an oblique MHD shock front, which is created by a shock wave sweeping a highly clumpy cloud; this is outlined in the literature and may be the main mechanism for filament formation in Cygnus X.
   \item Most prominent filamentary structures and high-mass cores are apparently located along the boundaries of HII regions or at the intersections of multiple HII regions. Filaments surrounding the HII regions have column-density profiles steeper toward the HII regions and flatter away from the HII regions, suggesting that the filaments are being compressed by the expanding HII regions. A comparison between the estimated lifetime of filaments and dynamical time scale of HII regions is compatible with speculation that the HII region expansion induces the filament formation, and the whole picture is analogous to the bubble-filament paradigm proposed by \citet{Inutsuka2015} and \citet{Pineda2023}.

\end{enumerate}

\section{Data availability}
The extracted cores and filaments are only available in electronic form at the CDS via anonymous ftp to cdsarc.u-strasbg.fr (130.79.128.5) or via \url{http://cdsweb.u-strasbg.fr/cgi-bin/qcat?J/A+A/}.
\begin{acknowledgements}

We thank Guangmao Liao for his helpful assistance and valuable discussions regarding the use of the \emph{getsf} algorithm. His guidance greatly contributed to the data analysis presented in this work. This work is supported by National Natural Science Foundation of China (NSFC) grants No. 12425304, U1731237, and National Key R$\&$D Program of China 2023YFA1608204 and No. 2022YFA1603100. We also acknowledge the support from the China Manned Space Project with No. CMS-CSST-2021-B06.

\end{acknowledgements}

\bibliography{Cygfil}{}         

\bibliographystyle{aa}

\begin{appendix}

\section{Comparision with \emph{DisPerSE} and \emph{FilFinder}}\label{subsec:Compared with DisPerSE/}

We also employ the \emph{FilFinder} and \emph{DisPerSE} algorithms to extract filament skeletons in a region $\sim$ 1/3 degree$^{2}$ (Fig. \ref{fig:Disperse}). For \emph{FilFinder}, we applied 1.2 $\times$ $10^{22}$ cm$^{-2}$ for global threshold and 0.4 pc for adaptive threshold. For \emph{DisPerSE}, we used a relative “persistence” threshold of 1 $\times$ 10$^{21}$ cm$^{-2}$. In Fig. \ref{fig:Disperse}, the red dashed box in panel (a), which highlights high column-density regions with prominent filaments, shows that \emph{getsf} identifies a single, well-defined, and continuous filament skeleton, whereas \emph{FilFinder} and \emph{DisPerSE} tend to produce multiple spurious closed loops. In contrast, the blue dashed box in panel (a), corresponding to low column-density regions, demonstrates that \emph{getsf} again recovers a complete filament skeleton, while \emph{FilFinder} and \emph{DisPerSE} extract only short or fragmented portions of the same filament. Although all three algorithms deliver broadly similar filamentary structures, \emph{getsf} behaves better than the other two in both high and low density regions. In addition, over the entire molecular cloud complex it is very difficult to find a parameter setup which is suitable for the whole column-density map to extract filaments with \emph{FilFinder} and \emph{DisPerSE}. Thus we finally choose \emph{getsf} in our work.

\renewcommand{\thefigure}{A1}

   \begin{figure*}[h!]
        \centering
        \includegraphics[width=0.3\textwidth]{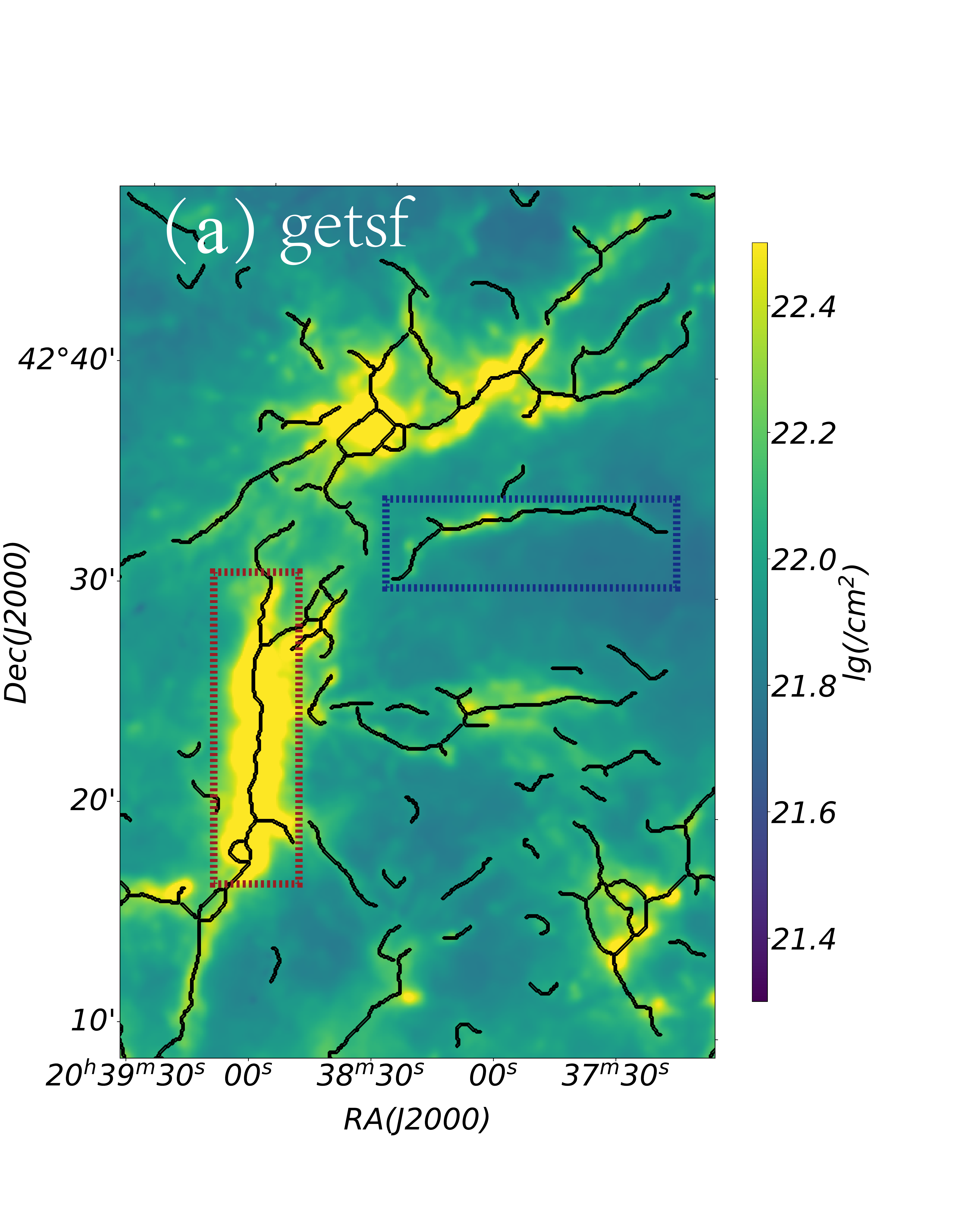}
        \includegraphics[width=0.3\textwidth]{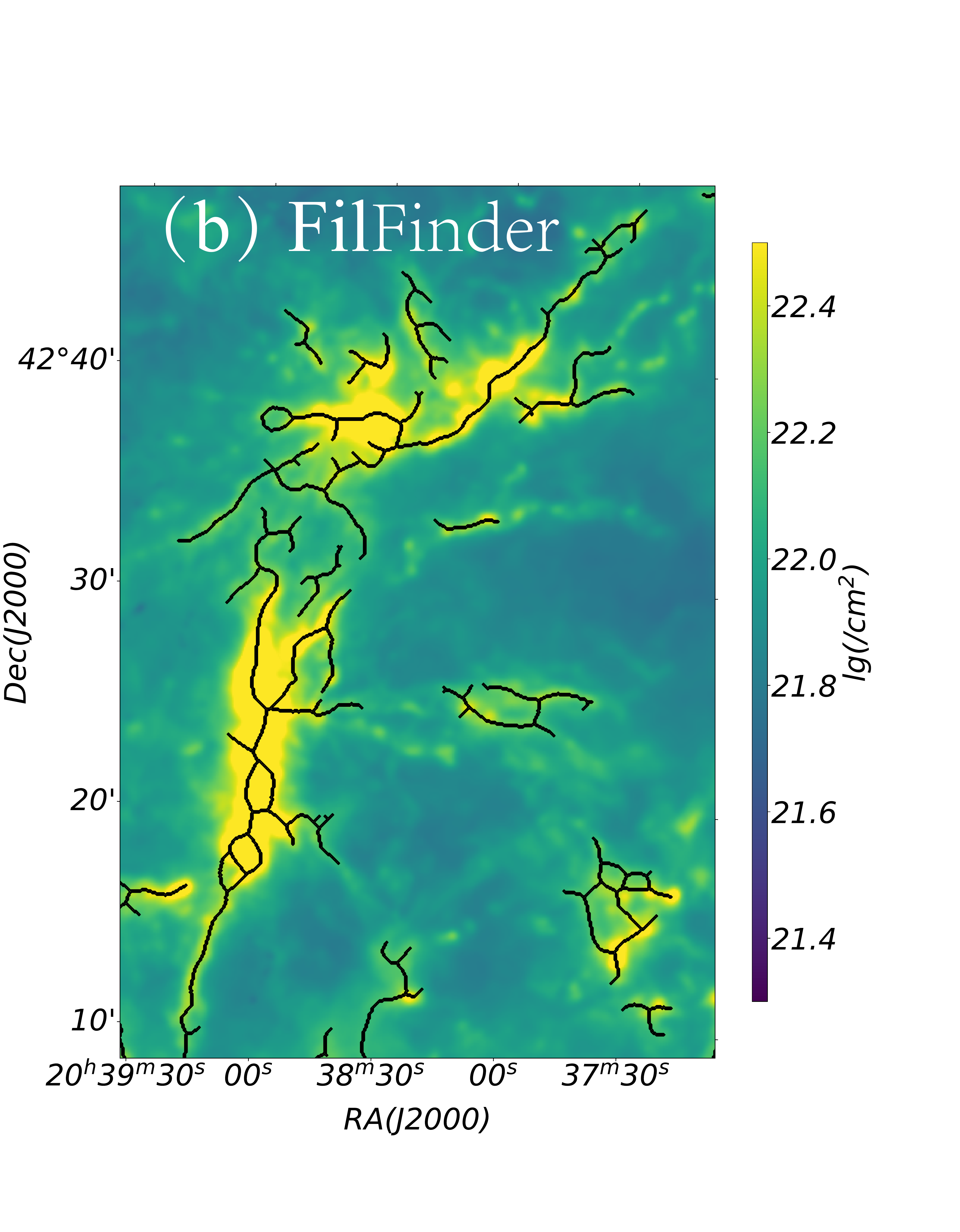}
        \includegraphics[width=0.3\textwidth]{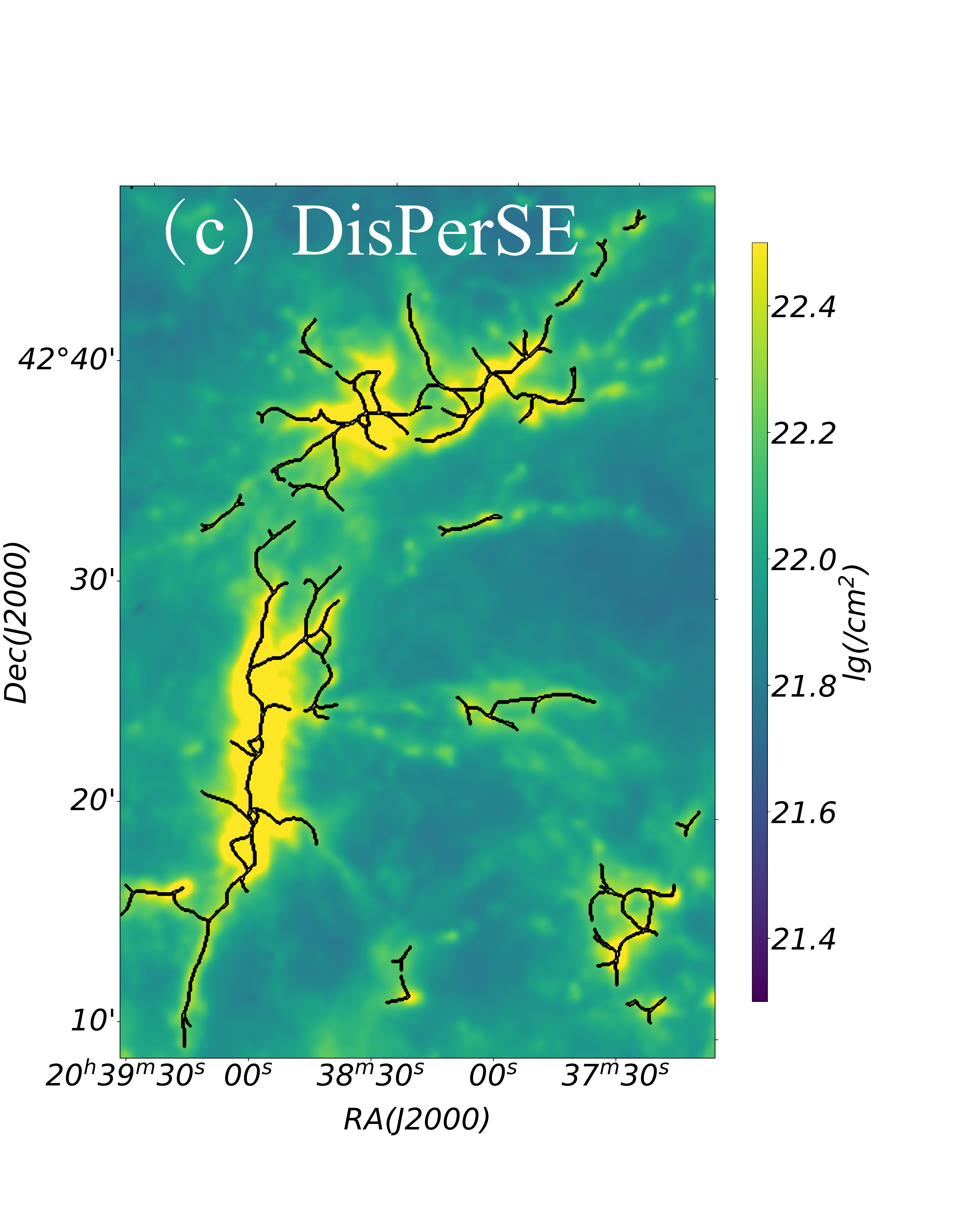}        
        \caption{Skeletons extracted with \emph{getsf}, \emph{FilFinder} and \emph{DisPerSE}. a) Skeletons extracted with \emph{getsf}; b) Skeletons extracted with \emph{FilFinder}; c) Skeletons extracted with \emph{DisPerSE}. Three algorithms show similar extraction results. However, the three algorithms still show clear differences in details. At high-density regions with prominent filaments (red dashed box), \emph{getsf} identifies a well-defined skeleton, whereas \emph{FilFinder} and \emph{DisPerSE} tend to produce two many closed loops. In low-density areas (blue dashed box), \emph{getsf} again extracts a complete filament skeleton, while \emph{FilFinder} and \emph{DisPerSE} can only deliver a small portion of the filament. Thus, although all three algorithms deliver broadly similar results, \emph{getsf} behaves better than the other two.}
        \label{fig:Disperse}
    \end{figure*}
\FloatBarrier

\section{ Five-step preprocessing sequence for width measurement} \label{subsec:The five-step preprocessing sequence in the width measurement/}

The filament width was measured following the method described in \citet{Suri2019}. Fig. \ref{fig:width} shows this five-step sequence in details.

\renewcommand{\thefigure}{B1}

    \begin{figure*}[h!]
    \centering
    \begin{tabular}{ccc}
        \includegraphics[width=0.25\textwidth]{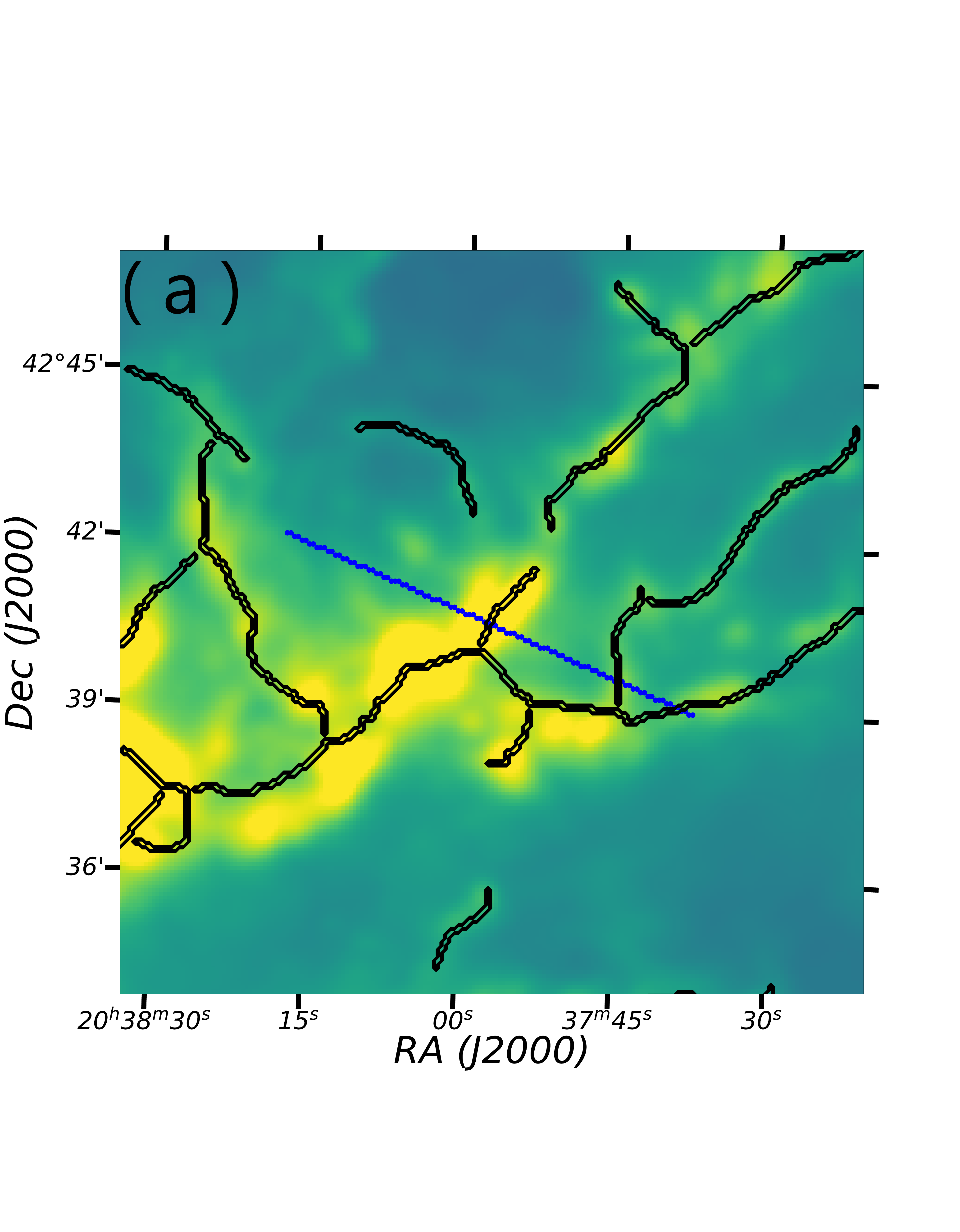}& 
        \includegraphics[width=0.3\textwidth]{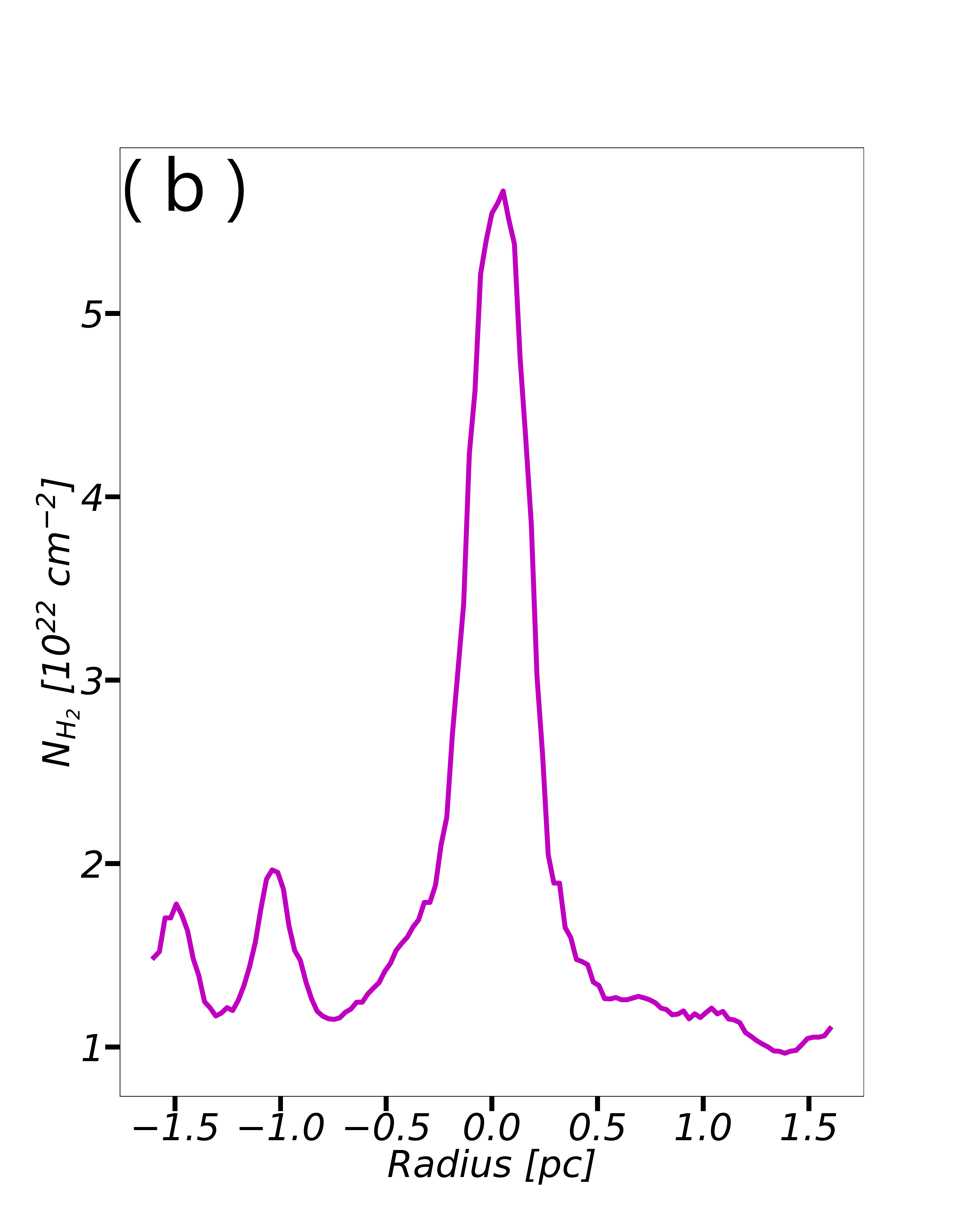}&
        \includegraphics[width=0.3\textwidth]{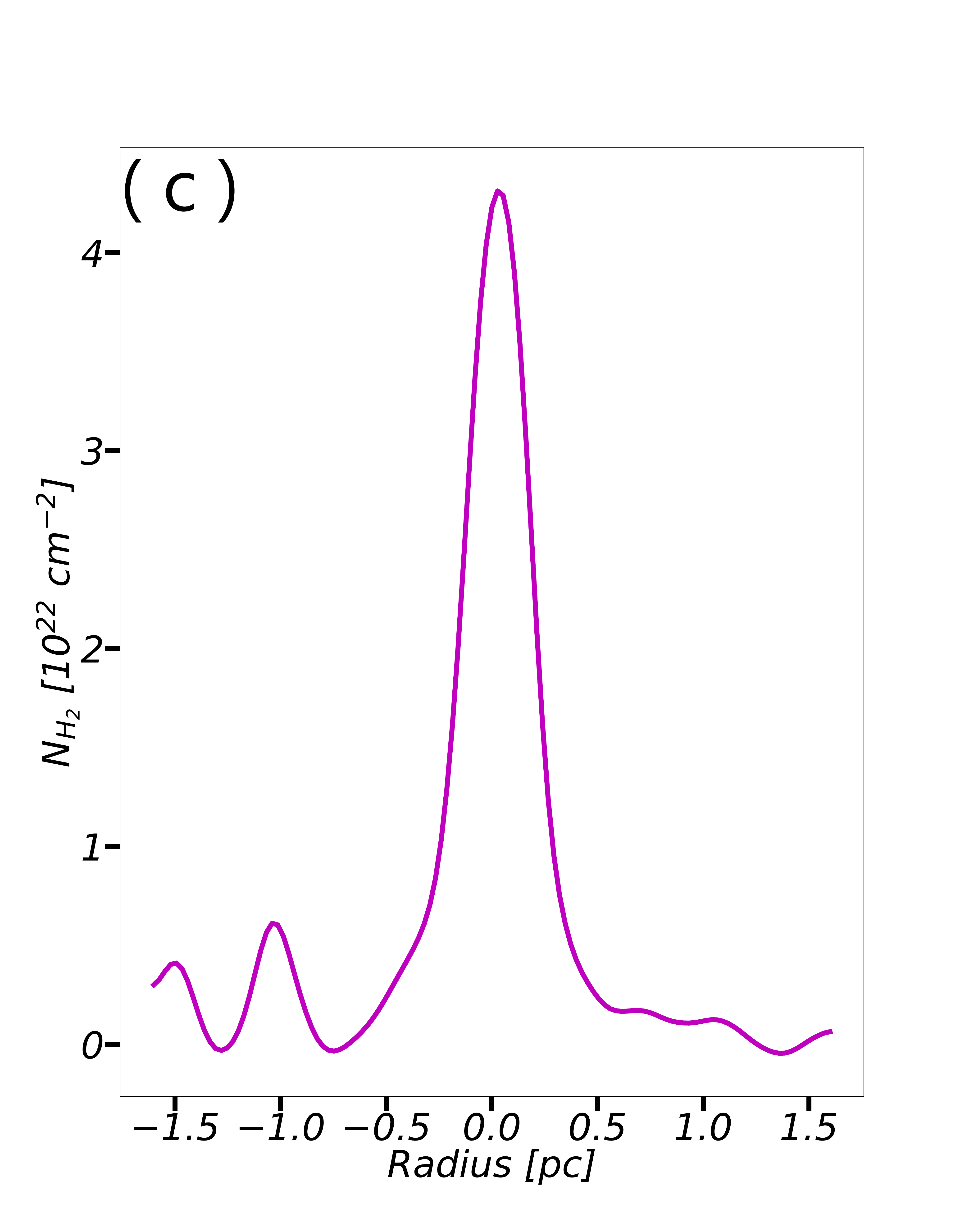}\\
        \includegraphics[width=0.3\textwidth]{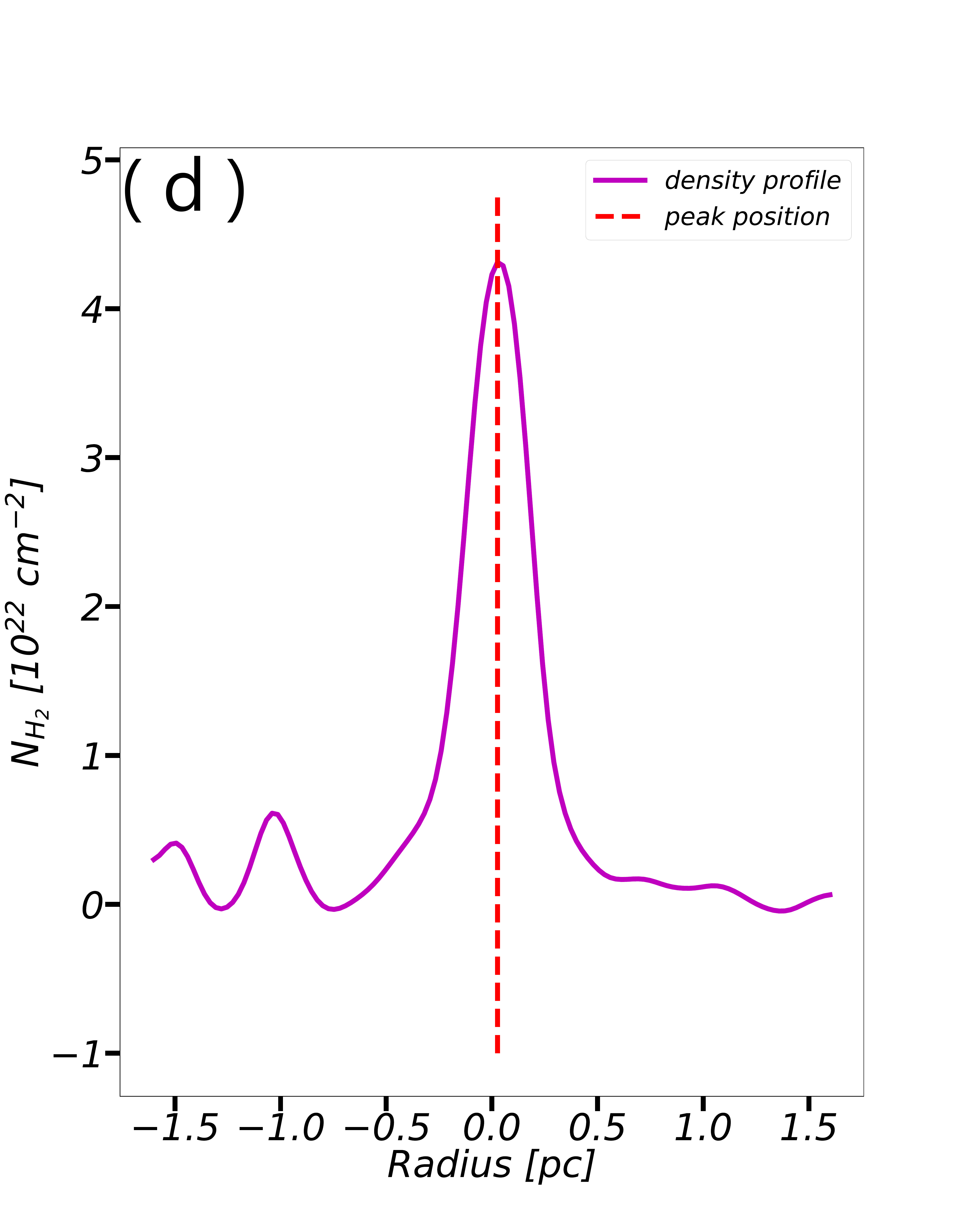}& 
        \includegraphics[width=0.3\textwidth]{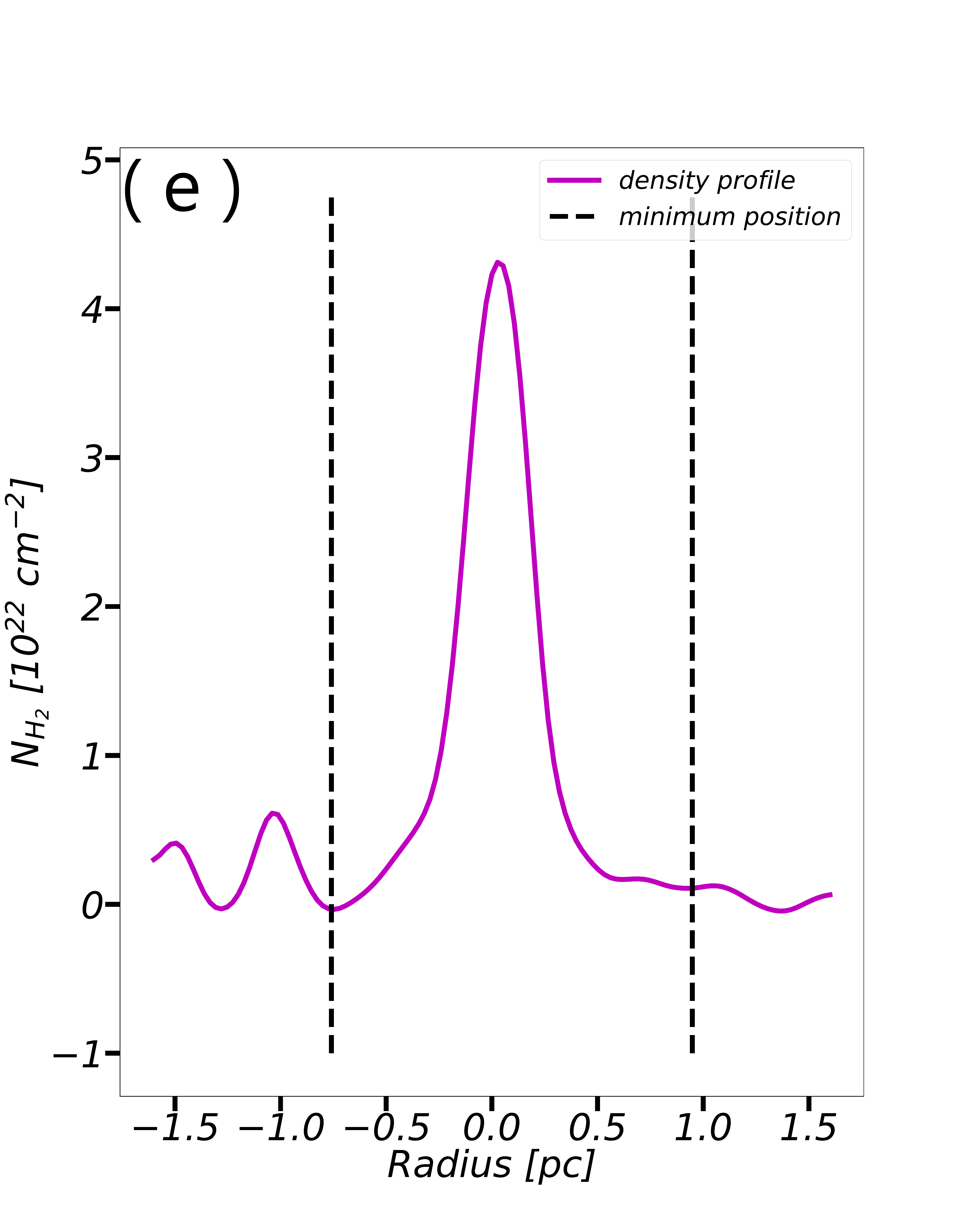}&
        \includegraphics[width=0.3\textwidth]{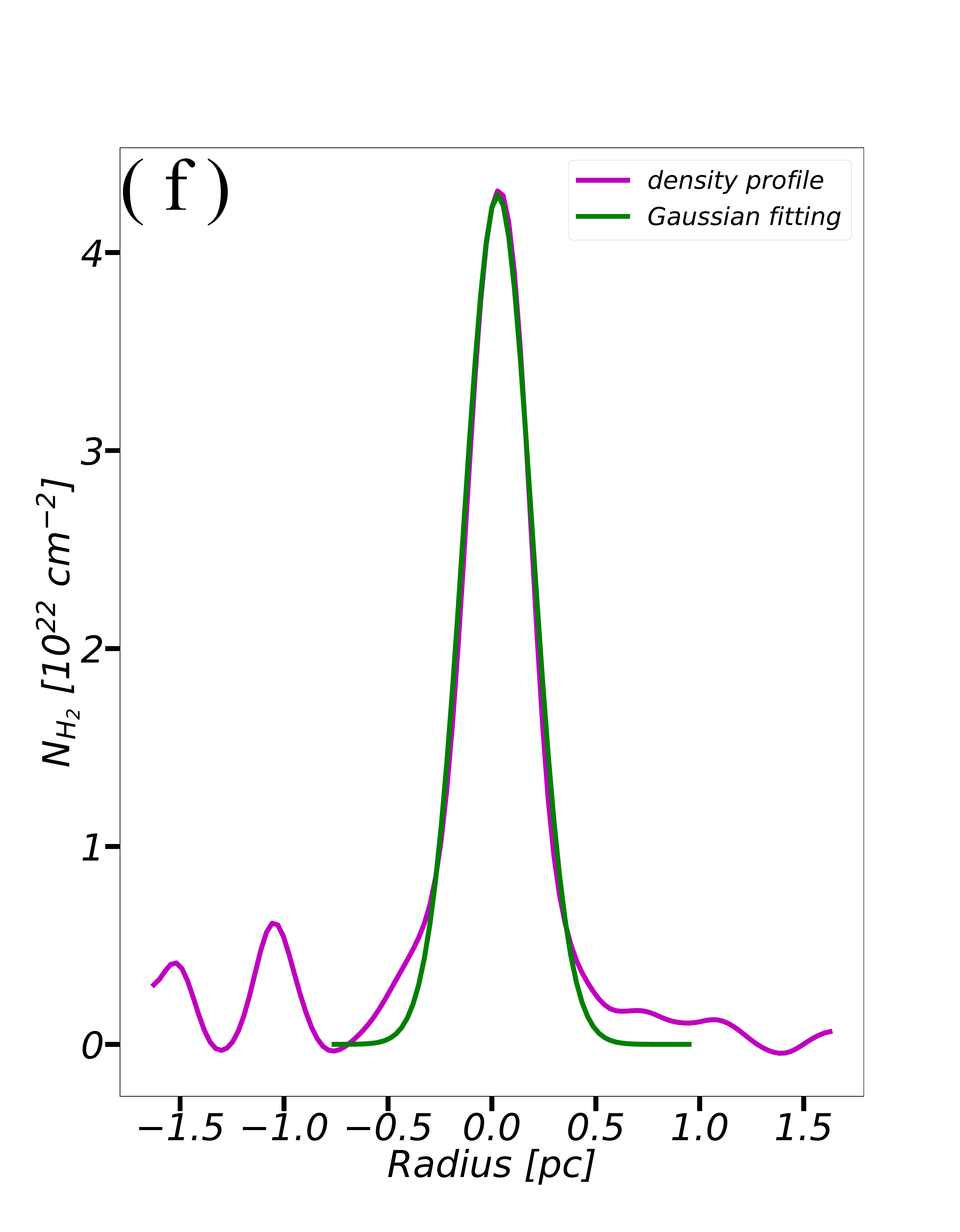}
    \end{tabular}
    \caption{Width calculation procedure: a) A zoom-in region cut from the column-density map with the extracted filament skeletons. The blue line represents a slice perpendicular to the skeleton. b) The original density distribution along the slice. c) The density distribution smoothed with a Gaussian kernel of our beam size (18.4") and then baseline subtraction. d) Identify the peak. e) Determine the minima nearest to the peak. f) Gaussian fitting within the constraints determined in the previous step.}
    \label{fig:width}
    \end{figure*}
\FloatBarrier

\section{Width fitting quality} \label{subsec:Impact of various models on the fitting qualities and width values/}

To determine the goodness of the width measurement based on the Gaussian fits of the column-density profiles described in Sec \ref{subsec:filament width} and Appendix \ref{subsec:The five-step preprocessing sequence in the width measurement/}, we adopt the coefficient $R^{2}$ as a unified evaluation criterion \citep{Carpenter1960}:
\begin{equation}
R^{2}=1-\frac{\Sigma(y_{\rm actual}-y_{\rm predict})^2}{\Sigma(y_{\rm actual}-y_{\rm mean})^2},
\end{equation}
where $y_{\rm actual}$ is the original data, $y_{\rm predict}$ is the fitted data, and $y_{\rm mean}$ is the average value of $y_{\rm actual}$. The width measurement is considered to improve as $R^{2}$ approaches 1. We show the $R^{2}$ distributions for the filament width measurements in three regions, which are selected to show varying levels of complexity in filamentary structures, in Fig. \ref{fig:gaussR2}.

\renewcommand{\thefigure}{C1}

    \begin{figure}[h!]
    \centering
    \includegraphics[width=0.6\columnwidth]{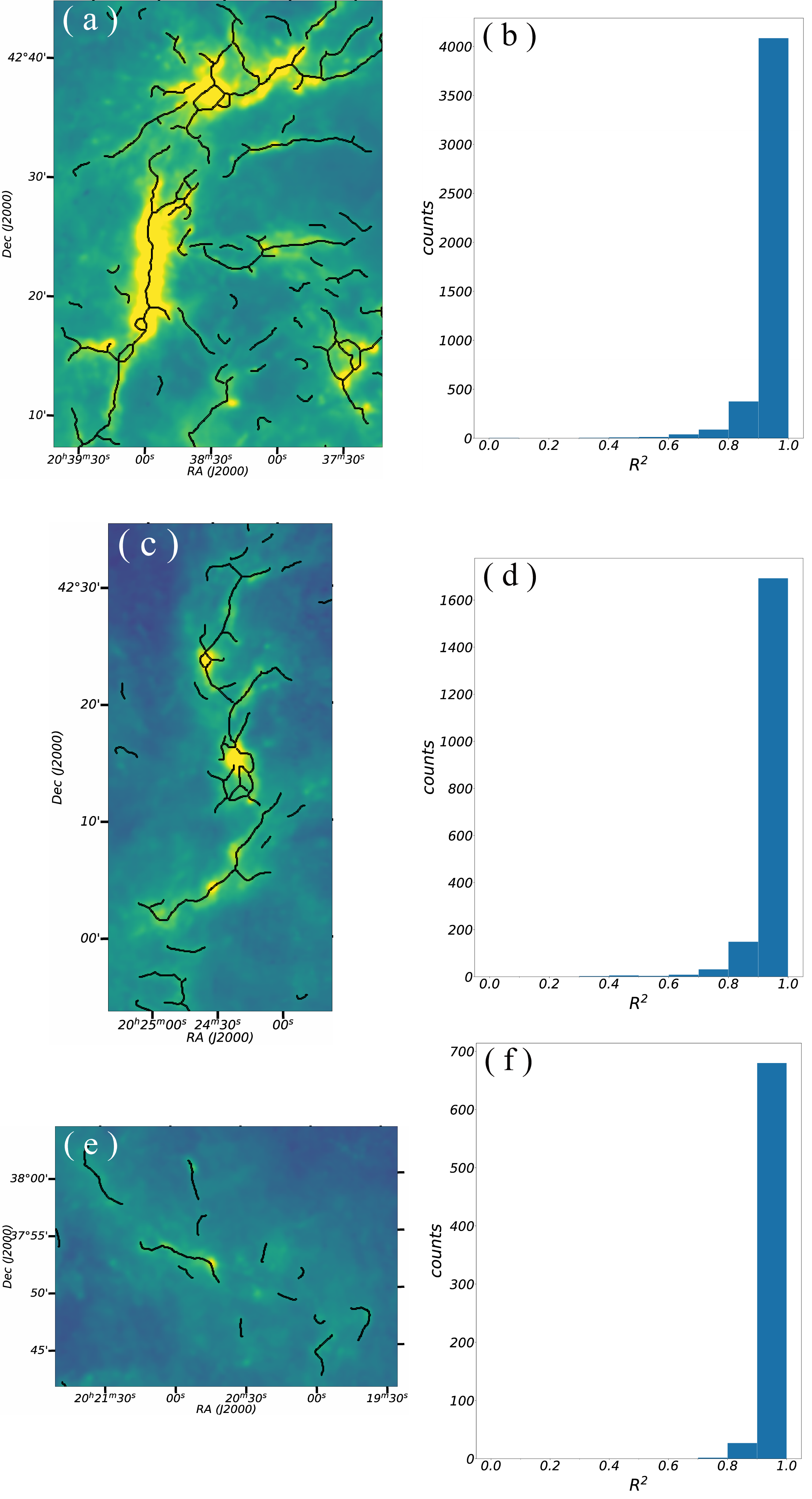}
    \caption{ The column-density maps with extracted filaments and the $R^{2}$ in the fitting of width measurements. a), c), e): column-density maps containing three types of filaments: filaments in a hub-filament system, the isolated filament and sparse filaments, respectively. b), d), f): $R^{2}$ derived from the fitting of width measurements in the regions shown in the left panels, respectively. All of the three $R^{2}$ distributions are mainly above 0.8.}
    \label{fig:gaussR2}
    \end{figure}
\FloatBarrier
\section{Filament extraction from the smoothed column-density map, and the orientation difference between the filaments and magnetic field} \label{subsec:orientation between filaments and magnetic field based on the mismatch resolution/}

To match the resolution of the B field map, we smooth our column-density map to the resolution of \emph{Planck} 353 GHz map ($5'$) and re-extract the skeletons on this map. We further exclude the filaments which are shorter than $5'$ or close to the boundaries within $5'$. The final extracted filaments are shown in Fig. \ref{fig:planckbone}.

We also check the orientation difference between the filaments extracted from the original (18.4" resolution) column-density map and the \emph{Planck} B field (see the left panel of Fig. \ref{fig:wrongorientation}). The distribution shows a perpendicular preference between the two entities.

To examine whether the orientation difference shown in the left panel of Fig. \ref{fig:wrongorientation} indicates a preferential perpendicular trend rather than a random distribution, we performed a Kolmogorov–Smirnov (K–S) test. The resulting probability (p-value) is 1.8 $\times$ 10$^{-33}$, which is far smaller than the significance level of 0.05, indicating that the null hypothesis of a random distribution can be statistically rejected. It should be noted, however, that the observed distribution of the orientation differences could be affected by the projection effect. To address this, we try to construct the distribution of the orientation differences in 3D, ${\theta}_{\rm 3D}$ from the observed distribution in 2D, ${\theta}_{\rm 2D}$. We assume that each pair of the filament and B field segments defines a plane having a random inclination angle with respect to the plane of sky. Note that in 3D, the angle $\alpha$ between two randomly oriented vectors is not uniformly distributed between 0 to 90 degree, but cos($\alpha$) has a uniform distribution from 0 to 1 \citep{Soler2013}. Thus we show the distribution of the corrected orientation differences in 3D as a function of $1-cos({\theta}_{\rm 3D})$ (the right panel of Figure \ref{fig:wrongorientation}). We see that after correcting for the projection effect, there is still a weak tendency of being perpendicular between the filaments and the B field in Cygnus X.

\renewcommand{\thefigure}{D1}

   \begin{figure}[h!]
        \centering
        \includegraphics[width=\columnwidth]{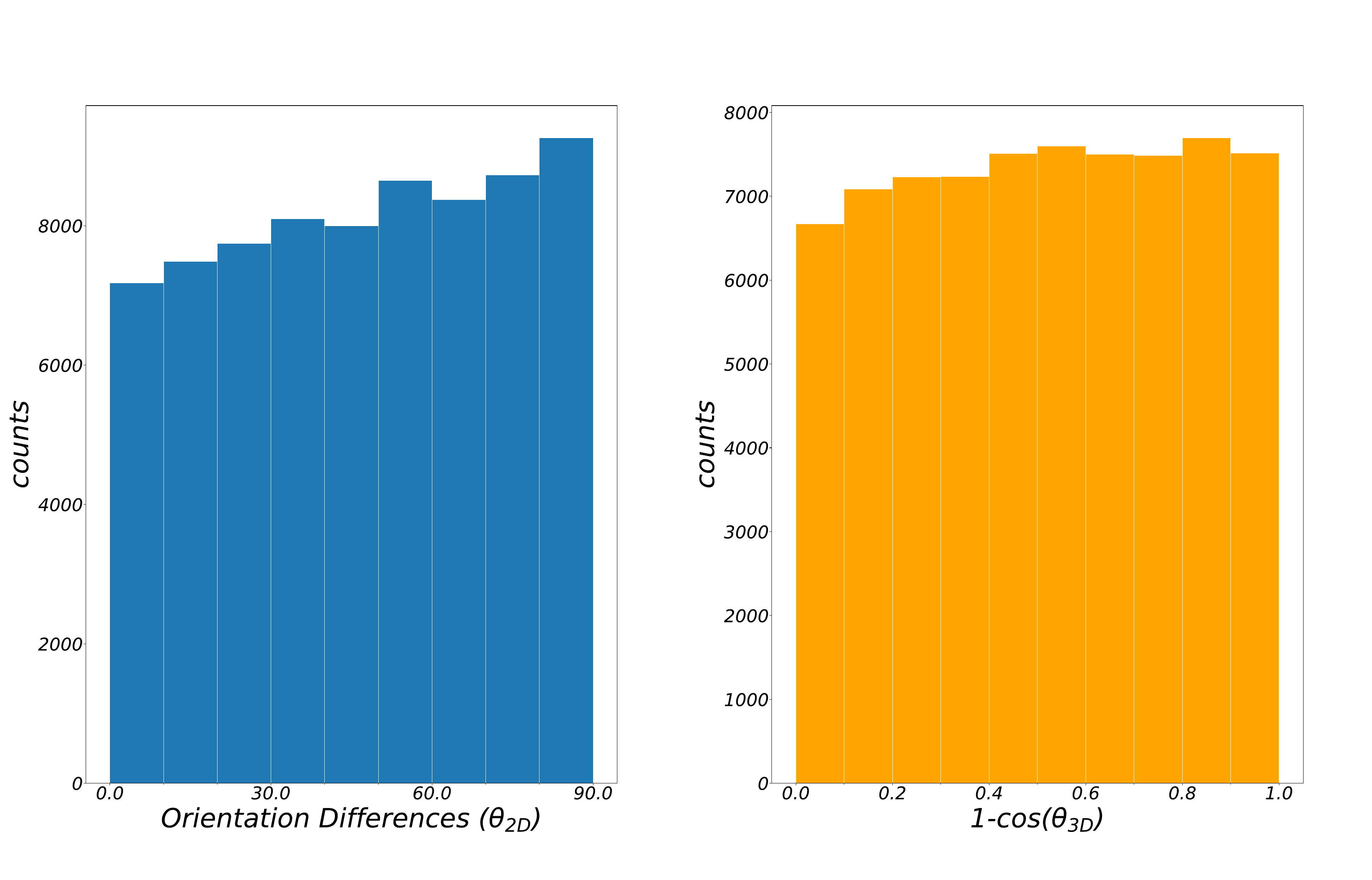}
        \caption{Left: the distribution of orientation differences between the B field (from \emph{Planck} 353 GHz dust polarization map) and the filament skeletons (extracted on our original column-density map with a resolution of 18.4$^{\arcsec}$). This distribution seems to show a perpendicular preference between the two entities. Right: the distribution of orientation differences in 3D shown as a function of $1-cos({\theta}_{\rm 3D})$, where ${\theta}_{\rm 3D}$ is the orientation difference angle in 3D as derived by correcting for a random inclination from ${\theta}_{\rm 2D}$, the observed orientation difference in 2D.}
        \label{fig:wrongorientation}
    \end{figure}

\section{The signal to noise ratios of the \emph{Planck} map} \label{subsec:The signal to noise of the Planck map/}

The Stokes I, Q and U from the \emph{Planck} data have their own uncertainties: $\delta I$, $\delta Q$ and $\delta U$. Considering the positive bias of polarized intensity PI and its associated uncertainty $\delta PI$, PI is derived following \citep{Vaillancourt2006}:

\begin{equation}
PI=\sqrt{Q^{2}+U^{2}-0.5(\delta Q^{2}+\delta U^{2})},
\end{equation}

and

\begin{equation}
\delta PI=\sqrt{\frac{(Q^{2} \delta Q^{2}+U^{2} \delta U^{2})}{(Q^{2}+U^{2})}},
\end{equation}

The debiased polarization percentage P and its uncertainty $\delta P$ are calculated as:

\begin{equation}
P=\frac{PI}{I}
\end{equation}

and

\begin{equation}
\delta P=\sqrt{(\frac{\delta {PI}^{2}}{I^{2}}+\frac{\delta I^{2}(Q^{2}+U^{2})}{I^{4}})}
\end{equation}

We compare the filament orientations and B-field orientations at 176 positions (Fig. \ref{fig:beam}). And the P/$\delta$P of 159 out of 176 ($\textgreater$ 90$\%$) positions exceeds 3.

\section{The relationship between the orientation differences and the densities} \label{subsec:The relationship between the orientation differences and the densities/}

Here we describe in detail the process of deriving the relationship between AM and the column densities. First, along the skeletons of each filament, we average the AMs and column densities over the pixels in each half \emph{Planck} beam ($\sim$ 37 pixels). This yields 150 data points. We then divide these 150 data points into 10 bins based on the logarithmic values of their column densities from low to high. For each bin, we again average the AMs and column densities, which finally gives us Fig. \ref{fig:bin10}. Results based on dividing the 150 data points into 20 and 30 bins are also shown in Fig. \ref{fig:bin2030}, which also shows that the filaments and B field tend to be parallel at low densities while change to be perpendicular at high densities.

\renewcommand{\thefigure}{F1}

    \begin{figure}[h!]
    \centering
    \includegraphics[width=\columnwidth]{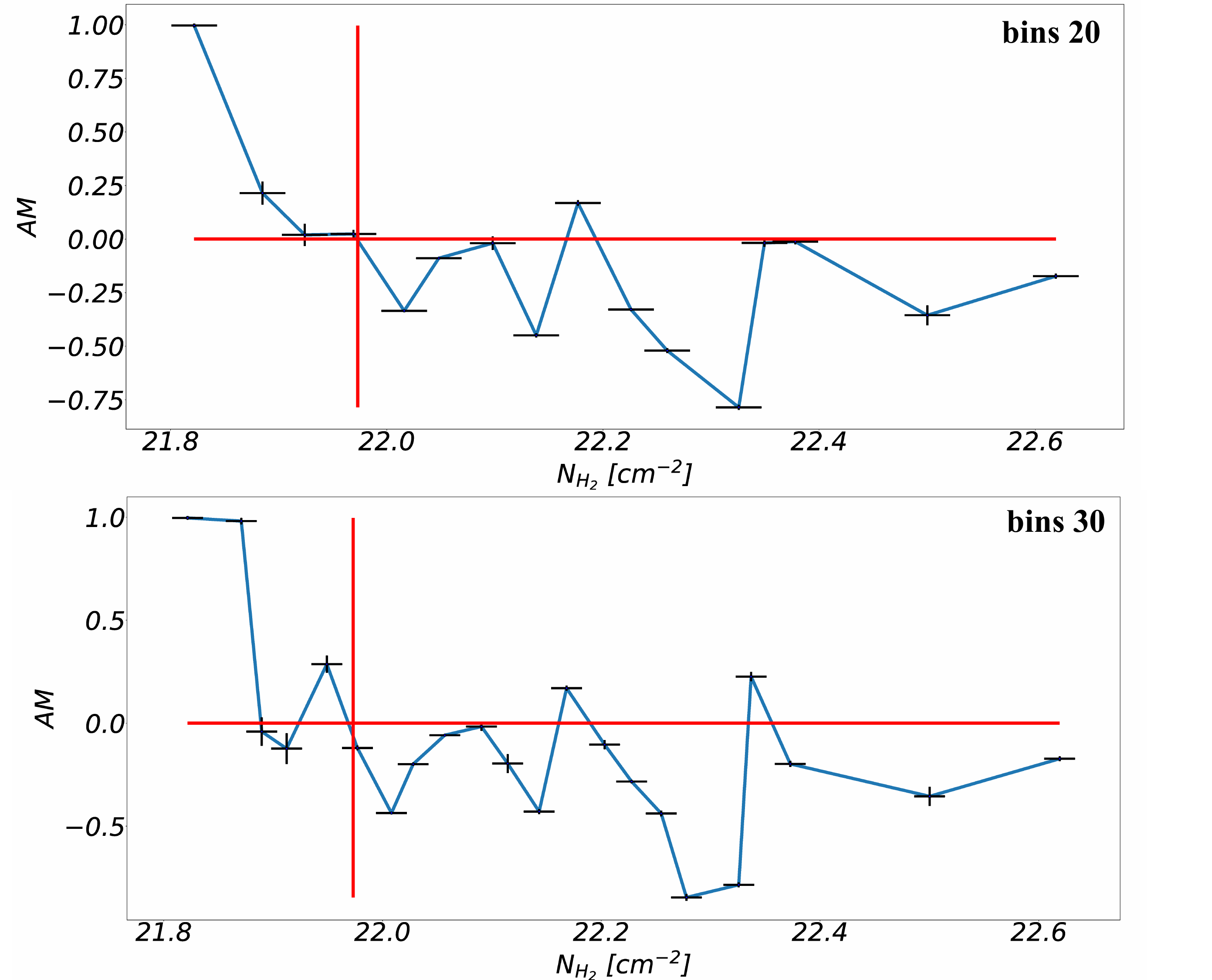}
    \caption{The AM versus column-density relation with 20 and 30 bins. Upper panel: the number of bins is 20. Bottom panel: the number of bins is 30. All lines and markers in this figure have the same meaning as Fig. \ref{fig:bin10}. Both of the two cases (bins 20 and bins 30) show a trend similar to that seen in Fig. \ref{fig:bin10}.}
    \label{fig:bin2030}
    \end{figure}

The uncertainty of filament orientations is calculated from the PCA reconstruction error. The reconstruction error here is defined as: $RE=||X-\hat{X}||^{2}$, where $X$ is the original data matrix and $\hat{X}$ is the reconstruction of $X$. Here, $\hat{X}=(X W)W^{\!\top}$, where $W$ is the main principal axis calculated by PCA. To calculate the uncertainty, we value the original data matrices as $X$ $+$ $\delta$$X$ ($\delta$$X$ uniformly drawn between -RE and RE), and repeat 100 times. We then apply PCA to each of these 100 new original data matrices and yield 100 new calculated filament orientations. The standard deviation of these new 100 orientations serves as the uncertainty of the filament orientation. The uncertainty of B field orientations is calculated with the formula Appendix \ref{subsec:The signal to noise of the Planck map/}. Finally, the total uncertainty in AM is then derived via standard error propagation, combining the uncertainties in both filament and magnetic‐field orientations.

\section{The difference between the filament skeleton and the density structure} \label{subsec:The difference between the filament and the density structure/}

The filament skeleton is extracted with special algorithm (like the \emph{getsf} used in this work) on the column-density map. However, the density structure is defined with the local column-density gradient $\nabla{\Sigma}$ \citep{PlanckCollaboration2016}. We note that, in the idealized case of an infinitely long cylindrical filament viewed side-on, these two entities would be perfectly perpendicularly aligned: the extracted filament skeleton would always be perpendicular to the local density structure across the filament crest. In reality, however, the situation is more complex (see Fig. \ref{fig:zoomindensitystructure}). As illustrated in the figure below, the two orientations tend to be nearly perpendicular in the vicinity of the high-density peak, whereas at lower column densities they appear to become more nearly parallel. Consequently, the orientation of the density structure does not provide a perfect proxy for the true filament direction in real clouds. This effect may contribute to the difference between the transition column-density, TN, derived in our work and that reported in \citet{PlanckCollaboration2016}.

\renewcommand{\thefigure}{G1}

    \begin{figure}[h!]
    \centering
    \includegraphics[width=\columnwidth]{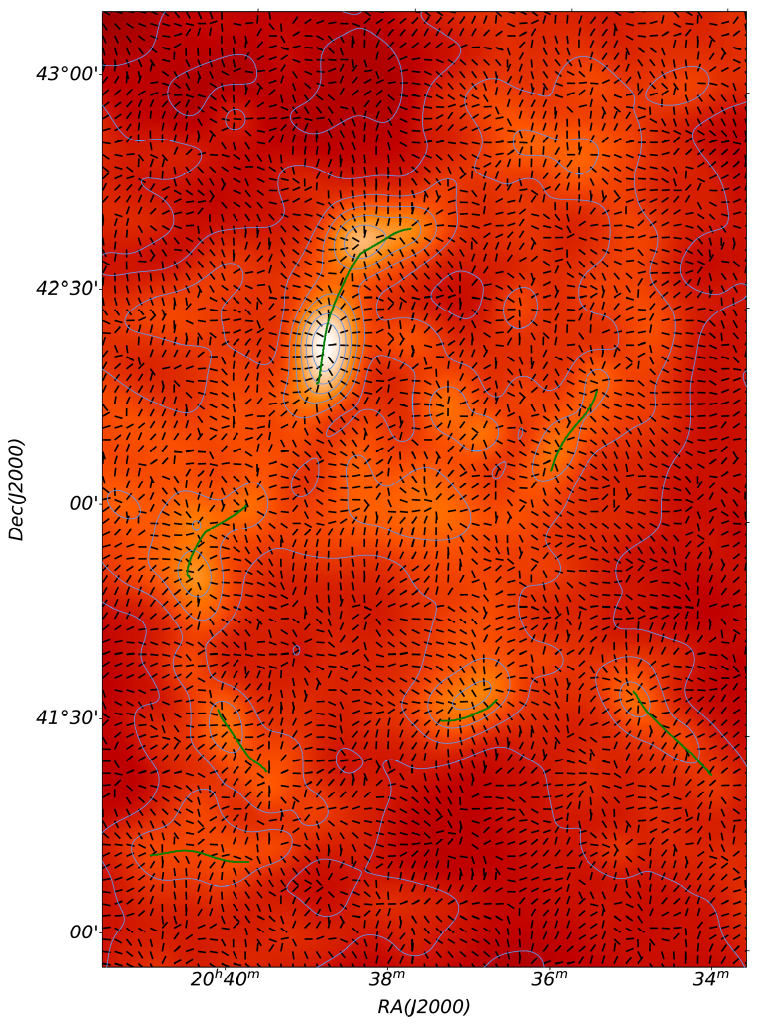}
    \caption{The filament skeleton extracted on the 5' resolution column-density map (green solid line), the density structure orientation (black segments) and the 5' column-density map (blue contour) in a zoomin region. The two orientations (the density structure and the filament skeleton) tend to be nearly perpendicular in the vicinity of the high-density peak, whereas at lower column densities they appear to become more nearly parallel.}
    \label{fig:zoomindensitystructure}
    \end{figure}

\end{appendix}
\end{document}